\documentclass[aps,prb,twocolumn,floatfix]{revtex4-2}

\usepackage{psfrag,epsfig,amsfonts,amssymb,amsmath,wasysym,bm}
\usepackage{bbold}
\usepackage{color}

\usepackage[inline]{enumitem} 

\renewcommand{\Re}{\operatorname{Re}}
\renewcommand{\Im}{\operatorname{Im}}

\newcommand{\kB}{k_{\mathrm B}} 

\newcommand{\RR}{{\mathbb R}}
\newcommand{\CC}{{\mathbb C}}
\newcommand{\NN}{{\mathbb N}}

\newcommand{\tr}{\operatorname{tr}}

\newcommand{\ord}{{\cal O}}

\newcommand{\lmat}{\left( \begin{matrix}}	
\newcommand{\rmat}{\end{matrix} \right)}	

\providecommand{\opnorm}[1]{\|#1\|_{\!\!\; op}}

\newcommand{\hh}{G}

\begin{document}

\title{Thermalization in a simple spin-chain model}

\author{Peter Reimann}
\author{Christian Eidecker-Dunkel}
\affiliation{Faculty of Physics, 
Bielefeld University, 
33615 Bielefeld, Germany}
\date{\today}

\begin{abstract}
We consider the common spin-1/2 XX-model in one dimension with 
open boundary conditions
and a large but finite number of spins.
The system is in thermal equilibrium at times $t<0$,
and is subject to a weak local perturbation (quantum quench) at $t=0$.
{Focusing mainly on single-spin perturbations and observables,}
we show that the system re-thermalizes for sufficiently large
times $t>0$ without invoking any unproven assumptions
besides the basic laws of quantum mechanics.
Moreover, the 
time-dependent relaxation behavior
is obtained in quantitative 
detail and is found to exhibit a wealth of 
interesting features.
\end{abstract}

\maketitle

\section{Introduction}
\label{s1}

Thermalization in isolated many-body systems is a well-established 
empirical observation in numerous experiments and numerical
investigations \cite{gog16,dal16,mor18,ued20}.
On the other hand,
obtaining rigorous analytical results for some given 
model system is still considered as a very daunting task 
{\cite{tas23,tas24a,tas24b,roo24}}
if one admits no unproven assumption, 
postulate, or hypothesis besides the basic laws 
of quantum mechanics 
(see also Sec.~\ref{s2} below).
Here, we analytically show thermalization for 
a particularly simple model, 
namely the so-called XX-spin chain, after a weak local quantum quench.
More precisely speaking, the system starts out in the canonical ensemble
(Gibbs state)
pertaining to some given pre-quench Hamiltonian, and then evolves under
the action of a slightly (locally) perturbed post-quench Hamiltonian.
{Focusing mainly on single-spin perturbations},
re-thermalization is shown in the sense that arbitrary 
single-spin
observables (and arbitrary sums thereof) approach the canonical
expectation values pertaining to the post-quench Hamiltonian
for the vast majority of all sufficiently late times
(after initial relaxation processes have died out).

Put differently, one arbitrary but fixed 
spin may be viewed as the 
subsystem of actual interest, 
and the entire rest of the chain
as its environment (thermal bath).
This subsystem is shown to (re-)thermalize after a small
{perturbation} in the presence of the thermal bath.

Moreover, not only the long-time behavior, but also the entire 
time-dependent relaxation is analytically obtained and 
discussed.
It exhibits a surprisingly rich structure, including some
key features which have also been observed 
(but not analytically explained)
before in many numerical explorations of more general models.

Technically, the main new ingredient is a rigorously provable
generalization of Onsager's regression hypothesis, recently
obtained in Ref.~\cite{rei24}.
Everything else in principle amounts to an application of
standard methods, albeit the 
detailed calculations are quite involved and 
to our knowledge not readily available in the literature.

\section{General Framework}
\label{s2}

Consider an isolated many-body quantum system in a 
(pure or mixed) initial state $\rho(0)$, evolving under the action 
of some model Hamiltonian $H$ according to von Neumann's equation
${i \dot \rho(t)}=[H,\rho(t)]$
into the state
\begin{eqnarray}
\rho(t) = e^{{-iHt}}\rho(0)\,e^{{iHt}}
\label{1}
\end{eqnarray}
{in units with}
\begin{eqnarray}
{\hbar = 1
\ .}
\label{1a}
\end{eqnarray}
The expectation value  of any observable 
(Hermitian operator) $A$ at time  $t$ 
is thus given by
\begin{eqnarray}
\langle A\rangle_t:=\tr\{\rho(t) A\}
\ .
\label{2}
\end{eqnarray}
Denoting the eigenvalues of $H$ by $E_n$ and 
the eigenvectors by $|n\rangle$,
one readily finds that 
\begin{eqnarray}
\langle A\rangle_t
=
\sum_{mn}
\rho_{mn}(0) A_{nm} e^{-{i(E_m-E_n)t}}
\ ,
\label{3}
\end{eqnarray}
where $\rho_{mn}(0):=\langle m  | \rho(0)|n\rangle$ and
$A_{nm}:=\langle n|A|m\rangle$.

As usual, in cases where we are interested in a system in contact with some
environmental bath, the Hamiltonian $H$ is supposed to describe
the (isolated) ``supersystem''  composed of the 
system
of interest  and the bath.
Moreover, only special choices of $A$ and $\rho(0)$ 
may then be of actual interest.

While {Eq.}~(\ref{3}) is still entirely general and exact, 
it is obvious that some additional knowledge about the 
quantities $E_n$, $\rho_{mn}(0)$, and $A_{nm}$ 
is indispensable to 
admit any 
physically relevant conclusions.

For instance, one may try to show that the system exhibits the
property of equilibration, meaning that the expectation values
in {Eq.}~(\ref{3}) stay very close to some constant value
(namely the long time-average of $\langle A\rangle_t$)
for the vast majority of all sufficiently late times $t$.
Rigorous proofs of such a behavior have 
been obtained for instance in 
Refs.~\cite{rei08,lin09,sho11,rei12,sho12,bal16}
under certain (quite weak) assumptions regarding 
$E_n$, $\rho_{mn}(0)$, and $A_{nm}$.
Most importantly,
the energies $E_n$ must not
exhibit extremely high degeneracies.
Similarly, extremely high degeneracies of the
energy gaps $E_n-E_m$ must be excluded.
Furthermore, the level populations $\rho_{nn}(0)$
(obeying $\rho_{nn}(0)\geq 0$ and $\sum_n\rho_{nn}(0)=1$) must be very 
small for all $n$, apart from possibly one exceptional $n$.
For systems with many degrees of freedom
there are quite convincing physical arguments 
why those assumptions may be expected to be
fulfilled in many cases, yet
a rigorous analytical verification for a given 
model Hamiltonian 
and initial state 
is extremely difficult.
Indeed, a rigorous demonstration of 
equilibration
without any such 
physically reasonable
but unproven extra assumptions 
has only been achieved very recently for some
quite special models 
and observables 
(while the initial states 
may 
be rather general), see Refs.~\cite{tas23,tas24a,tas24b,roo24}.

As a next step, one may take equilibration for granted 
and try to show that the system exhibits the property of
thermalization, meaning that the long-time average
of $\langle A\rangle_t$ is (approximately) equal to
the thermal expectation value 
$\tr\{\rho_{mc}A\}$, where $\rho_{mc}$ is the
pertinent 
microcanonical ensemble.
This property can be verified for instance under the
extra assumption that the so-called eigenstate 
thermalization hypothesis (ETH) is fulfilled 
\cite{neu29,deu91,sre94,rig08}
(see also \cite{gog16,dal16,mor18,ued20}).
However, until now the ETH has the status of a 
hypothesis, which to the best of our knowledge has
not {been} rigorously derived for any 
(non-trivial)
model Hamiltonian $H$ and observable $A$.
Alternative approaches to rigorously demonstrate 
thermalization for 
some given Hamiltonian and initial state
have been devised for instance in 
Refs.~\cite{far17,dab22}, however still utilizing 
(among others) some of the above 
mentioned unproven assumptions in the 
context of equilibration.

More precisely speaking, a given system 
(and initial state) is commonly 
regarded as exhibiting
thermalization if every physically relevant observable 
does so \cite{gog16,dal16,mor18,ued20}.
Our first remark is that for 
any 
non-thermal
initial state 
$\rho(0)$ 
there exist some Hermitian operators $A$ which do not thermalize.
Typically, these $A$ 
involve projectors onto 
eigenstates of $H$ 
and are usually viewed as being
physically unrealistic
(not experimentally feasible),
at least for (reasonable) model systems with many 
degrees of freedom, as we consider them here.
Therefore, the most commonly adopted 
standpoint is that only so-called local 
(i.e., short-ranged and few-body)
operators $A$ 
must be 
taken into account
\cite{gog16,dal16,mor18,ued20}.
We also adopt this viewpoint in our present work.
Note that thermalization then immediately follows 
also for arbitrary sums of local operators.
Moreover, macroscopic 
observables 
of physical relevance
are supposed to be always given by 
such sums of local operators,
and are thus covered as well.

It is a {long-known and} 
commonly appreciated fact  that 
integrable systems generally do not exhibit 
thermalization after a 
global quench, {see for example 
Refs.~\cite{gog16,dal16,mor18,ued20,bar70}.}
 On the other hand, thermalization has been analytically
proven in Refs.~\cite{tas23,tas24a,tas24b,roo24} 
for systems which are integrable 
and initial conditions which include global 
quenches as special cases.
However, only some subset of observables 
has actually been 
addressed in these works, and there still must 
exist some other (physically relevant) 
observables, which do not thermalize
\cite{foot0}.

Also in our present work, the considered models 
are integrable, and hence are known 
to generally avoid thermalization after a {\em global} quench
(see also Sec.~\ref{s42}). 
This is the reason why we will only address {\em local} 
quenches.
Furthermore,
%
the number of degeneracies and
degenerate energy gaps (see above) will
turn out to increase exponentially with the 
system size. Hence, all the above-mentioned 
previous results
\cite{rei08,lin09,sho11,rei12,sho12,bal16}
regarding the issue of equilibration 
will be useless in our present case.

On the other hand, in the definition of thermalization
we will replace the microcanonical ensemble 
$\rho_{mc}$ by the corresponding 
canonical ensemble 
\begin{eqnarray}
\rho_{can} & := & e^{-\beta H}/\tr\{e^{-\beta H} \}
\ ,
\label{4}
\\
\beta & := & 1/\kB T
\ ,
\label{5}
\end{eqnarray}
where $T$ is the temperature and $\kB$ Boltzmann's 
constant, and where $\beta$ is (as usual)
chosen such that $\tr\{\rho_{mc}H\}=\tr\{\rho_{can}H\}$.
In other words, we take for granted the
equivalence of ensembles in the sense that
$\tr\{\rho_{mc}A\}=\tr\{\rho_{can}A\}$ for
all physically relevant observables $A$.
A more rigorous justification of this equivalence 
is provided for instance in \cite{tou15,bra15,tas18,kuw20}.
For the rest, 
this issue is not the actual main subject of 
our present paper.
From an alternative viewpoint
one could 
in fact
employ $\rho_{can}$ instead of $\rho_{mc}$ already
in the very definition of thermalization.
The main reason is that in cases where 
the equivalence of the ensembles 
might be violated, 
it is not obvious which of them is the 
``right'' one anyway (without invoking some 
unproven hypothesis such as Jaynes principle).

For further general remarks and comparisons with previous
works we also refer to our Conclusions in Sec.~\ref{s6}.

\section{Setup}
\label{s3}

We focus on the XX-spin-chain model with open boundary conditions
and Hamiltonian
\begin{eqnarray}
H = 
- J \sum_{l=1}^{L-1}  \, (s^x_{l+1}s^x_l  + s^y_{l+1}s^y_l)
- \hh \sum_{l=1}^{L} \, s_l^z
\, ,
\label{6}
\end{eqnarray}
where $s^{a}_l$
with $a\in\{x,y,z\}$  
are spin-1/2 operators at the chain sites
$l\in\{1,...,L\}$,
$J$ quantifies the nearest neighbor interactions,
and $\hh$ the coupling to a homogeneous 
transverse field.
Furthermore, we restrict ourselves to the non-trivial case $J\not  = 0$,
and we work in units 
such that the operators $2s_l^a$ can be represented as
Pauli matrices in the eigenbasis of $s_l^z$ {(see also Eq.~(\ref{1a})).}

It is well-known that this Hamiltonian can be analytically 
diagonalized by mapping it via a Jordan-Wigner transformation 
to an equivalent model of free (spinless) fermions.
The details of this transformation are provided in 
Appendix \ref{appA}.
Accordingly, such models are sometimes classified as 
noninteracting integrable systems in the recent literature
\cite{spo18}.
Furthermore, the number of degeneracies and
degenerate energy gaps of the Hamiltonian (\ref{6})
can be shown to grow 
exponentially with the system size $L$, 
see Appendix \ref{appA5}.
Yet another 
general property of
this Hamiltonian is that it commutes
with $S^z:=\sum_{l=1}^L s_l^z$,
i.e., $S^z$ is a conserved quantity.

At time $t=0$ the system is assumed to be 
in an initial state of the form
\begin{eqnarray}
\rho(0) & := & e^{-\beta H_{\! g}}/\tr\{e^{-\beta H_{\! g}}\}
\, ,
\label{7}
\end{eqnarray}
where 
\begin{eqnarray}
H_{\! g}:=H - g V
\label{8}
\end{eqnarray}
is a slight modification of the original Hamiltonian 
{from Eq.~}(\ref{6}) 
with a local perturbation operator $V$ 
and  a 
{coupling} parameter $g$.
While $\rho(0)$ in {Eq.}~(\ref{7}) thus amounts to a 
thermal equilibrium state (canonical ensemble) with respect to
the modified Hamiltonian $H_{\! g}$, it is a non-equilibrium 
initial state with respect to the actual Hamiltonian $H$ 
of the system we are considering, which then evolves
for $t>0$ 
according to {Eq.}~(\ref{1}),
yielding
time-dependent expectation values
as defined in {Eq.}~(\ref{2}).
Equivalently, these expectation values 
can be rewritten as
\begin{eqnarray}
\langle A\rangle_t & = & \tr\{\rho(0) A(t)\} 
\, ,
\label{9}
\end{eqnarray}
where
\begin{eqnarray}
A(t)
& := &
e^{{iH t}} A e^{{-iHt}}
\label{10}
\end{eqnarray}
is the observable at time $t$ in the
Heisenberg picture.

As discussed in Sec.~\ref{s2}, the problem of thermalization
then amounts to the question whether $\langle A\rangle_t$
stays sufficiently close to the
thermal equilibrium value
\begin{eqnarray}
A_{\rm th} :=\tr\{\rho_{can} A\}
\label{11}
\end{eqnarray}
for the vast majority of all sufficiently late times $t$,
where $\rho_{can}$ is the canonical ensemble from {Eq.}~(\ref{4}).
We remark that, in principle, the values of $\beta$ in {Eqs.}~(\ref{4})
and (\ref{7}) could still be different.
But since we focus on 
local perturbations in
{Eq.}~(\ref{8}) it is reasonable to expect, and will 
even be rigorously verified later on,
that the values of $\beta$ can indeed be 
chosen identical in our present case.

Similarly as in {Eqs.}~(\ref{9})-(\ref{11}),
the temporal correlation
(also called, among others, dynamic or 
two-point correlation function) of 
two Hermitian operators 
$V$ and $A$ at thermal equilibrium 
is defined as
\begin{eqnarray}
C_{\! V\!\!A}(t)
& := & 
\langle V\! A (t) \rangle_{\rm th}
- V_{\rm th} A_{\rm th}
\ .
\label{12}
\end{eqnarray}
While $V$ and $A$ in this definition are still arbitrary, we 
henceforth always
focus on the situation that $V$ is the perturbation
operator appearing in the initial condition (\ref{7}) via {Eq.}~(\ref{8}),
and $A$ some local observable.
(As already said in Sec.~\ref{s2}, the extension to sums of
such observables is immediate.)
Under this premise we then can utilize  
the following result from \cite{rei24},
\begin{eqnarray}
 \langle A\rangle_t - \langle A\rangle _{\rm th}
 & = &  
g \beta \sum_{k=0}^\infty  \frac{(i {\beta})^k}{(k+1)!}\,  \frac{d^k}{dt^k} 
C_{V\!A}(t)
\ ,
\label{13}
\end{eqnarray}
connecting the time-dependent expectation values of a system 
out of equilibrium
with the temporal correlations at thermal equilibrium.

Our first remark is that {Eq.}~(\ref{13}) is essentially
a linear response kind of approximation in the sense 
that higher order corrections in $g\beta$ have been 
neglected on the right-hand side.
{More precisely speaking, it has been shown in Ref. \cite{rei24} that
the neglected terms 
can be upper bounded (in modulus) by 
$\opnorm{A} (\epsilon^2+\ord(\epsilon^3))$,
where $\opnorm{A}$ is the operator norm of $A$
(largest eigenvalue in modulus).
Thus, we always may add a trivial constant to $A$ 
in order to minimize $\opnorm{A}$, and then 
utilize a trivial multiplicative factor so that
$\opnorm{A}=1$.
Furthermore,
\begin{eqnarray}
\epsilon := g\beta \opnorm{V} f_H(\beta)
\ ,
\label{13a}
\end{eqnarray}
where $f_H(\beta)$ 
depends on $\beta$ and on the
properties of the unperturbed system Hamiltonian $H$.
For our specific model from {Eq.}~(\ref{6}),
the function $f_H(\beta)$ turns out to be of order unity
at least as long as the parameters $\beta J$ and $\beta G$ 
are 
of order unity
(or smaller).
We thus can conclude that the approximation (\ref{13}) 
will be very good if one of the two quantities $\beta$ 
and $g\opnorm{V}$ is sufficiently small.
In particular, one of them may not be small 
if the other is all the smaller.
Moreover, under the extra condition that 
$\beta J$ and $\beta G$ are at most 
of order unity, it is sufficient that the product
$g\beta \opnorm{V}$ is small.
Note that for extensive perturbations $V$, 
the operator norm  $\opnorm{V}$ generically
diverges as the system size increases,
hence our approximation (\ref{13}) becomes 
questionable
(it may still be valid, but there is not yet a proof).
This is another reason (besides the one already mentioned 
above Eq.~(\ref{4})) 
why we mainly focus on local perturbations in our present work.
We finally emphasize that all these considerations 
regarding the accuracy of Eq.~(\ref{13})
are valid for arbitrary  times $t$ 
and chain-lengths $L$ of the unperturbed system in Eq.~(\ref{6}).}

{Our second remark is that the correlations in Eq.~(\ref{12}) are in general
complex-valued functions of $t$, and likewise for every single 
summand in Eq.~(\ref{13}).
However, the entire sum can be shown to be real-valued \cite{rei24},
and hence every summand can be replaced by its real part.
By keeping only the real part of the first summand 
(which equals Eq.~(\ref{12})),}
one recovers the so-called Onsager regression hypothesis.
While this hypothesis is well-known to become quantitatively 
incorrect  beyond the classical limit, the main achievement
of \cite{rei24} is a rigorous derivation of the amended 
relation (\ref{13}). 
Employing this relation plays a key role
in our present work, since we will be able to 
exactly evaluate $C_{V\!A}(t)$  for a large class of
operators $V$ and $A$, while we did not succeed 
(even for our simple model {in Eq.}~(\ref{6}))
to directly evaluate the left-hand side of {Eq.}~(\ref{13}).

\section{Main examples}
\label{s4}

As our first specific examples we consider 
perturbations $V$ and observables $A$ of the form
\begin{eqnarray}
V=s_\nu^z\, , \ \ A=s_\alpha^z
\label{14}
\end{eqnarray}
with arbitrary $\nu, \alpha \in\{1,...,L\}$.
{The systems's initial state 
is thus given by the canonical ensemble from Eqs.~(\ref{7}) 
and (\ref{8}), where $H$ is the XX-model from Eq.~(\ref{6}) 
and $V=s_\nu^z$, and this initial state
then evolves in time 
according to Eq.~(\ref{1}).}
As detailed in Appendix~\ref{appC}
(which in turn is based on Appendices 
\ref{appA} and \ref{appB}), it 
{is 
possible} to
evaluate the corresponding temporal correlations 
{in Eq.~}(\ref{12}) as well as
the infinite sum on the right-hand side of {Eq.}~(\ref{13})
in closed analytical form 
without any further approximation,
resulting in
\begin{eqnarray}
\langle s_\alpha^z \rangle_t - \langle s_\alpha^z\rangle _{\rm th}
& = & 
g \beta \!
\sum_{m,n=1}^L \tilde S_m^\nu \tilde S_n^\nu \tilde S_m^\alpha \tilde S_n^\alpha  f_{mn}
\, e_{mn}(t)
\ ,
\ \ \ \ \ \ 
\label{15}
\\
\tilde S_k^l & := & \sqrt{\mbox{$\frac{2}{L+1}$}}\, \sin(\pi k l /(L+1))
\ ,
\label{16}
\\
 f_{mn}
& := & 
\frac{\tanh(\beta \tilde E_m/2)-\tanh(\beta \tilde E_n/2)}{2\beta(\tilde E_m-\tilde E_n)}
\ ,
\label{17}
\\
e_{mn}(t)
& := &
e^{{-i(\tilde E_{m}-\tilde E_{n})t}}
\ ,
\label{18}
\\
\tilde E_k & := & - J \cos(\pi k/(L+1)) - \hh
\ .
\label{19}
\end{eqnarray}
In other words, the only approximations in this 
result are those 
{discussed around Eq.~(\ref{13a}) 
(complemented by $\opnorm{A}=\opnorm{V}=1/2$ according to Eq.~(\ref{14}),
and working in units as specified below Eq.~(\ref{6})).}
The purpose of the tilde symbols is to better distinguish,
for instance, $\tilde E_k$ from $E_n$ (see above Eq.~(\ref{3})),
or $\tilde S_k^l$ from $s_l^a$ 
and $S^z$ (see below Eq.~(\ref{6})).

One readily verifies that the right-hand side of {Eq.}~(\ref{15})
is a real-valued function of $t$.
Therefore, only the real part of
$e_{mn}(t)$ in {Eq.}~(\ref{18}) actually contributes, and
the result {in Eq.}~(\ref{15}) is time-inversion invariant.

\begin{figure}
\hspace*{-0.8cm}
\includegraphics[scale=0.95]{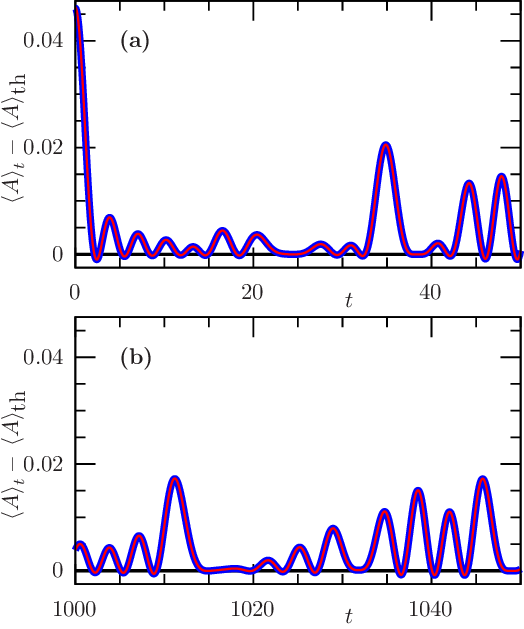}
\caption{
Blue: Numerically exact 
{results for the expectation values} 
$\langle A\rangle_t - \langle A\rangle _{\rm th}$,
{employing}
the XX-model {from Eq.}~(\ref{6}) with $L=16$,
{$J=1$}, and $\hh =0$, 
{the perturbation $V$ and observable $A$ 
from Eq.}~(\ref{14}) 
with $\nu=\alpha=L/2$,
{and the} 
initial condition {from Eqs.~(\ref{7}) and (\ref{8})} with $\beta=1$ and $g=0.2$.
Red: {Corresponding} analytical approximation {from Eq.}~(\ref{15}).
(a): Initial relaxation behavior for $t\in[0,50]$.
(b): Long-time behavior for $t\in[1000, 1050]$.
The differences between the red and 
blue curves are at most about $10^{-4}$,
and hence 
not visible on the scale of this plot.
}
\label{fig1}
\end{figure}

As detailed in Appendix \ref{appA}, the quantities $\tilde S_k^l$ 
in {Eq.}~(\ref{16})
are the elements of a unitary matrix (see Eq.~(\ref{a78})), 
i.e., they satisfy  the orthonormality relations
\begin{eqnarray}
\sum_{\nu=1}^L \tilde S_m^\nu \tilde S_n^\nu = \delta_{mn}
\ ,\ \ 
\sum_{n=1}^L \tilde S_n^\nu \tilde S_n^\alpha = \delta_{\nu\alpha}
\ .
\label{20}
\end{eqnarray}
[Alternatively, one also may derive these relations directly from {Eq.}~(\ref{16}).]
Concerning the quantities $ f_{mn}$ in {Eq.}~(\ref{17})
one can furthermore show that
\begin{eqnarray}
0< f_{mn} \leq  1/4 \ \ \mbox{for all $m,n$,}
\label{21}
\end{eqnarray}
as detailed in Appendix \ref{appD} (see Eq.~(\ref{d4}) therein).

A first remarkable feature of {Eq.}~(\ref{15}) is the symmetry
with respect to 
the indices $\nu$ and $\alpha$, and thus with respect to
the perturbation $V$ and the observable $A$
from {Eq.}~(\ref{14}).
Second, the transverse field strength $\hh$ from
{Eq.}~(\ref{6}) only enters into {Eq.}~(\ref{15}) via {Eq.}~(\ref{17}), 
while {Eq.}~(\ref{18}) is independent of $\hh$ 
according to {Eq.}~(\ref{19}).
Third,
the number of summands in {Eq.}~(\ref{15}) grows
only quadratically with the system size $L$,
while such sums (for instance in {Eq.}~(\ref{3}))
are usually expected to grow exponentially 
with the system size.
We speculate that this
might possibly be
due to our special (integrable noninteracting) 
model (see below Eq.~(\ref{6})),
or due to the fact that only linear 
order terms in $g\beta$ have been kept
(see {above Eq.~(\ref{13a})}).
Our last remark is that 
by summing over $\alpha$ in {Eq.}~(\ref{15}) and
exploiting {Eq.}~(\ref{20}) one readily can infer that
$S^z:=\sum_{\alpha=1}^L s_\alpha^z$
is indeed a conserved quantity
(see above Eq.~(\ref{7})).

\begin{figure}
\hspace*{-0.8cm}
\includegraphics[scale=0.95]{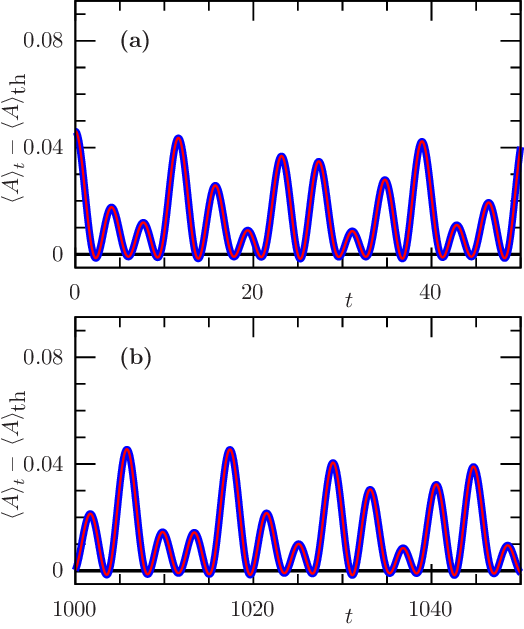}
\caption{
Same as in Fig.~\ref{fig1} but for a smaller system with $L=4$.
}
\label{fig2}
\end{figure}

A further general symmetry property is based on the 
trivial identity
\begin{eqnarray}
\sum_{n=1}^L c_n = \sum_{n=1}^L c_{L+1-n}
\label{22}
\end{eqnarray}
for arbitrary 
summands $c_n$.
We thus may replace all indices $n$ in {Eq.}~(\ref{15})
by $L+1-n$, and similarly for $m$.
According to {Eq.}~(\ref{16}), this means that 
$\tilde S_m^\nu$ is replaced by $(-1)^{\nu+1}\tilde S_m^\nu$
and analogously for all the other factors in {Eq.}~(\ref{15}).
Denoting the right-hand  side of {Eq.}~(\ref{15}) as
a function $F(\beta,J,\hh)$ of the three basic model 
parameters $\beta$, $J$, and $\hh $,
one thus 
finds after some elementary calculations that
 $F(\beta,J,\hh )=F(\beta,J,-\hh )$.
Recalling that only the real part of {Eq.}~(\ref{18}) 
actually counts (see below Eq.~(\ref{19}))
one furthermore can infer from {Eq.}~(\ref{15}) that
$F(\beta,-J,-\hh )=F(\beta,J,\hh )$.
Combining both findings yields
$F(\beta,-J,\hh )=F(\beta,J,\hh )$.
Finally, $F(-\beta,J,\hh )=-F(\beta,J,\hh )$
is immediately evident from {Eq.}~(\ref{15}).
Altogether, we thus can conclude that {Eq.}~(\ref{15}) must be an odd 
function of $\beta$, and an even function of $J$ and of $\hh $.

\begin{figure}
\hspace*{-0.8cm}
\includegraphics[scale=0.95]{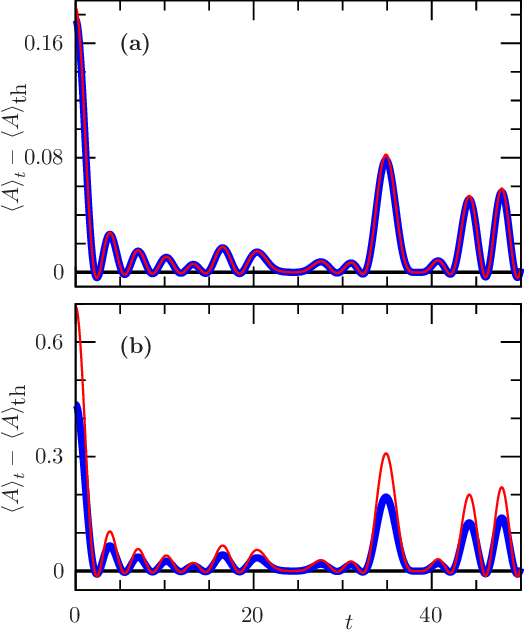}
\caption{
Same as in Fig.~\ref{fig1} (a) but for stronger perturbations
with $g=0.8$ in (a) and $g=3.0$ in (b).
The differences between the 
{blue and red}
curves 
are still 
{quite small} 
in (a), but clearly visible in (b).
}
\label{fig3}
\end{figure}

For the rest, the 
approximation (\ref{15}) is still quite complicated from 
a purely analytical viewpoint, but can be easily evaluated
numerically even for rather large chain-lengths $L$.
On the other hand, a direct, numerically exact evaluation of the
quantity on the left-hand side of {Eq.}~(\ref{15})
is practically only feasible by diagonalizing 
the full many-body Hamiltonian from {Eq.}~(\ref{6}), 
and is thus restricted to relatively small $L$-values.
Fig.~\ref{fig1} exemplifies such 
{numerically} exact results
 {(blue curves)}
together with the
(numerically evaluated) approximation from Eq.~(\ref{15})
 {(red curves).}
The close agreement of the two curves indicates
that the 
approximation is very accurate.
Moreover, the agreement is very good not only 
with respect to the initial relaxation behavior in (a), 
but also for the very late times in (b), 
as predicted below Eq.~(\ref{13a}).
Likewise, Fig.~\ref{fig2} illustrates the prediction 
below Eq.~(\ref{13a}) that the approximation works 
{equally} well for {a short chain with $L=4$
as for the longer chain with $L=16$ in Fig.~\ref{fig1}.}

{While the differences between the exact (blue)
and approximative (red) curves are hardly visible
in Figs.~\ref{fig1} and \ref{fig2}, 
larger values of $g$ have been chosen in
Fig.~\ref{fig3} in order to better visualize}
the higher order corrections in $g$, which have been 
neglected in the approximation (\ref{15}),
{see also the discussion around Eq.~(\ref{13a}).
Incidentally, one can show by symmetry arguments 
\cite{foot5}
that the left-hand side of Eq.~(\ref{15}) 
must be an odd function of the parameter $g$ 
provided that $G=0$.
Hence, the neglected terms on the right-hand 
side of the approximation (\ref{15}) 
must be (at least) of order $g^3$.
Fig.~\ref{fig3} 
shows a direct comparison between a
numerically exact evaluation of the quantity on the 
left-hand side of Eq.~(\ref{15}) (blue) and its approximation
as given by the right-hand side of {Eq.}~(\ref{15}) (red).
A closer inspection of Fig.~\ref{fig3} shows that
the differences between the two curves
are indeed in good agreement with this prediction 
that they should grow as $g^3$. }

\begin{figure}
\hspace*{-0.8cm}
\includegraphics[scale=0.95]{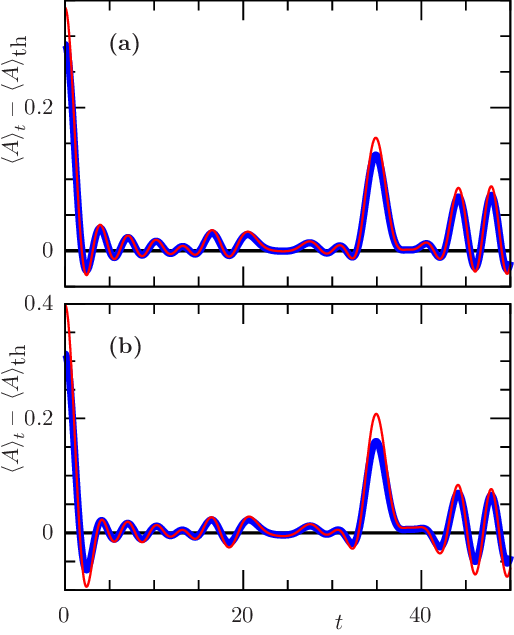}
\caption{
Same as in 
Fig.~\ref{fig3} (a)
but for larger $\beta$-values (smaller temperatures, 
see also {Eq.}~(\ref{5}) 
and {main text}),
namely $\beta=2$ in (a) and $\beta=4$ in (b).
}
\label{fig4}
\end{figure}

{Finally,}
Fig.~\ref{fig4} illustrates the analogous
finite-$\beta$ effects.
{Our first remark is that
we adopted in Fig.~\ref{fig4} the same, relatively large 
perturbation strength $g=0.8$ as in Fig.~\ref{fig3}~(a)
in order to obtain clearly visible differences between 
the blue and red curves.
Our second remark is that the curves in Fig.~\ref{fig4}~(b) 
were found to hardly change any more
upon further increasing $\beta$, i.e., we are 
already close to the zero temperature 
asymptotics for our present, still relatively 
small system with $L=16$.
Our last remark is that}
the relative differences between 
the red and blue curves {both} in Figs.~\ref{fig3} and 
\ref{fig4} seem to be 
approximately independent of $t$, but we did not succeed 
to come up with an analytical explanation 
of this numerical observation.

We believe that further 
numerical illustrations 
that {the approximation} 
(\ref{15}) indeed works very well also
for other values of $\hh$ in {Eq.}~(\ref{6})
and of $\nu$ and $\alpha$ in {Eq.}~(\ref{14})
may not be indispensable at this point.
Henceforth, we will thus take 
{Eq.}~(\ref{15}) 
for granted,
and
we will focus on a more detailed
discussion of the complicated expression on
the right-hand side.
Furthermore, in our subsequent plots of
{Eq.}~(\ref{15}) we can and will 
remove the dependence on $g$ by dividing both sides by $g$. 
(Incidentally, the dependence of {Eq.}~(\ref{15}) on $\beta$ is far less trivial,
see also Fig.~\ref{fig4}).
We finally remark that also the thermal expectation values 
$\langle s_\alpha^z\rangle _{\rm th}$ appearing in {Eq.}~(\ref{15})
can be analytically evaluated (without any approximation)
and discussed analogously to the right-hand side 
of {Eq.}~(\ref{15}),  as detailed in Appendix~\ref{appE}.

\subsection{Proof of thermalization}
\label{s41}

{This section is devoted to the derivation of thermalization 
as one of the key results of our paper.
We proceed as follows.
Taking the approximation (\ref{15}) for granted, we will
show in a first step that its long-time average
becomes arbitrarily small for sufficiently
large system sizes $L$ (thermodynamic limit).
In a second step we show that an analogous property
also applies to the corresponding temporal mean 
square fluctuations.
Finally, we 
explain why
these two findings imply the property of thermalization 
as specified in detail
in Sec.~\ref{s2}.}

To begin with, we rewrite {Eq.}~(\ref{15}) as
\begin{eqnarray}
\langle s_\alpha^z \rangle_t - \langle s_\alpha^z\rangle _{\rm th}
& = & 
\sum_{m,n=1}^L K_{mn}^{\nu\alpha} \, e^{i E_{nm} t}
\ ,
\label{23}
\end{eqnarray}
where 
\begin{eqnarray}
K_{mn}^{\nu\alpha}
& := &
g\beta
\tilde S_m^\nu \tilde S_n^\nu \tilde S_m^\alpha \tilde S_n^\alpha  f_{mn}
\ ,
\label{24}
\\
E_{nm}
& := &
{\tilde E_n-\tilde E_m}
\ .
\label{25}
\end{eqnarray}
From {Eq.}~(\ref{19}) one can infer that
$\tilde E_n\not=\tilde E_m$ for all $n,m\in\{1,...,L\}$ with $n\not=m$
(assuming $J\not=0$, see below Eq.~(\ref{6})).
Denoting the long-time average by an overline symbol, 
we thus can conclude that
\begin{eqnarray}
\overline{e^{i E_{nm} t}}=\delta_{nm}
\ .
\label{26}
\end{eqnarray}
Together with {Eqs.}~(\ref{23})-(\ref{25}) it follows that
\begin{eqnarray}
& &  \overline{\langle s_\alpha^z \rangle_t} - \langle s_\alpha^z\rangle _{\rm th}
=
g \beta \sum_{n=1}^L (\tilde S_n^\nu \tilde S_n^\alpha)^2  f_{nn}
\ .
\label{27}
\end{eqnarray}
Exploiting {Eqs.}~(\ref{16}) and (\ref{21}) we can conclude that
$0\leq  (\tilde S_n^\nu \tilde S_n^\alpha)^2  f_{nn}\leq 1/L^2$
and thus
\begin{eqnarray}
0\leq \overline{\langle s_\alpha^z \rangle_t} - \langle s_\alpha^z\rangle _{\rm th} \leq g\beta /L
\ \ \mbox{if $g\beta\geq 0$.}
\label{28}
\end{eqnarray}
Hence, the long-time average $\overline{\langle s_\alpha^z \rangle_t}$ must
approach the thermal equilibrium value $\langle s_\alpha^z\rangle _{\rm th}$
from above as the system size $L$ increases.
Analogous conclusions apply if $g\beta < 0$.

Next we turn to the temporal variance {(or mean square fluctuations)}
\begin{eqnarray}
\sigma^2:=\overline{\left(\langle s_\alpha^z \rangle_t -\overline{\langle s_\alpha^z \rangle_t}\right)^2}
\ .
\label{29}
\end{eqnarray}
Observing that $\langle s_\alpha^z \rangle_t -\overline{\langle s_\alpha^z \rangle_t}$
in {Eq.}~(\ref{29})
is equal to the difference between {Eq.}~(\ref{23}) and {Eq.}~(\ref{27}), it follows that
\begin{eqnarray}
\sigma^2
=
\overline{\Big(\sum_{m\not=n} K_{mn}^{\nu\alpha} \, e^{i E_{nm} t}\Big)^2}
\leq
\overline{\Big(\sum_{mn} K_{mn}^{\nu\alpha} \, e^{i E_{nm} t}\Big)^2}
\ ,
\ \ \ \ 
\label{30}
\end{eqnarray}
where the first sum runs over all $m,n\in\{1,...,L\}$ with $m\not=n$
and the second sum over all $m,n\in\{1,...,L\}$.
Finally, this implies
\begin{eqnarray}
\sigma^2
\leq
\sum_{mn} \sum_{jk} K_{mn}^{\nu\alpha}K_{jk}^{\nu\alpha} \, \overline{e^{i (E_{nm}-E_{jk}) t}}
\ .
\label{31}
\end{eqnarray}
Similarly as below Eq.~(\ref{27}) one can show that
$|K_{mn}^{\nu\alpha}|\leq | g \beta|  /L^2$, 
and with {Eq.}~(\ref{31}) that
\begin{eqnarray}
\sigma^2
& \leq &
(g\beta)^2 \, \gamma/L
\ ,
\label{32}
\\
\gamma
& := &
\frac{1}{L^3}
\sum_{mnj} \sum_{k} \left|\overline{e^{i (E_{nm}-E_{jk}) t}}\right|
\ .
\label{33}
\end{eqnarray}
Similarly as in {Eq.}~(\ref{26}), the summands in {Eq.}~(\ref{33}) are unity
whenever $E_{nm}=E_{jk}$ and zero otherwise.
According to {Eq.}~(\ref{25}), the 
latter criterion
is equivalent to 
\begin{eqnarray}
\tilde E_m-\tilde E_n+\tilde E_j=\tilde E_k
\ .
\label{34}
\end{eqnarray}
Let us temporarily consider $m,n,j$ as arbitrary but fixed. 
Since $\tilde E_k$ is a strictly monotonic function of $k$
according to {Eq.}~(\ref{19}), it follows that there is at most one
$k$ for which {Eq.}~(\ref{34}) is fulfilled.
As a consequence, the innermost sum over $k$ in
{Eq.}~(\ref{33}) can be upper bounded by unity, 
and thus the remaining triple sum over $n,m,j$ by $L^3$,
implying
\begin{eqnarray}
0 \leq \gamma \leq 1
\ .
\label{35}
\end{eqnarray}
It follows that $\sigma^2$ in {Eq.}~(\ref{32})
approaches zero as the system size
$L$ increases. 
In view of {Eq.}~(\ref{29}) we can conclude that $\langle s_\alpha^z \rangle_t$
must be close to $\overline{\langle s_\alpha^z \rangle_t}$
for the vast majority of all times $t$.
In turn, $\overline{\langle s_\alpha^z \rangle_t}$ approaches the 
thermal equilibrium value $\langle s_\alpha^z\rangle _{\rm th}$ 
for large $L$, see below Eq.~(\ref{28}).
Altogether, the system thus exhibits thermalization
(at least as far as the observables $A=s_\alpha^z$ from {Eq.}~(\ref{14})
are concerned, see also Secs.~\ref{s2} and \ref{s5}).
{A more detailed discussion of this issue is postponed to the next subsection.}

Incidentally, it seems reasonable to expect that for any given
pair $m,n$ the number of pairs $j,k$ with the property
$E_{nm}=E_{jk}$ is non-extensive in the system size $L$.
This implies that $\gamma$ in {Eq.}~(\ref{33}) decreases
as $1/L$ and thus $\sigma^2$ in {Eq.}~(\ref{32}) as $1/L^2$.
However, a rigorous derivation of such a stronger bound
(compared to {Eq.}~(\ref{35}))
is not obvious, and therefore not further pursued here.

\begin{figure}
\hspace*{-0.8cm}
\includegraphics[scale=0.95]{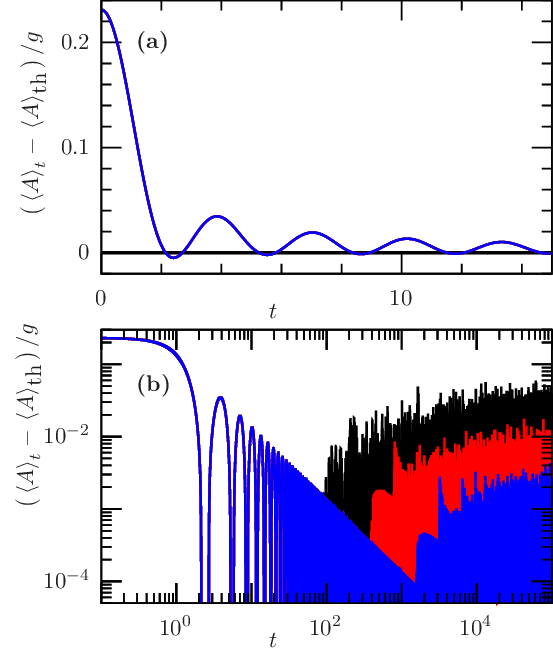}
\caption{
(a) Numerical evaluation of Eq.~(\ref{15}) 
divided by $g$ for 
$L=100$ (black), $L=400$ (red), $L=1600$ (blue).
Everything else as in Fig.~\ref{fig1}.
(b) Same data but now plotted double-logarithmically.
Up to $t\approx 100$ the black curves are covered by the blue ones,
and likewise for the red curves up to $t\approx 400$.
}
\label{fig5}
\end{figure}

\subsubsection{Discussion}
\label{s411}

Fig.~\ref{fig5} exemplifies these predictions in quantitative detail.
In particular, 
Fig.~\ref{fig5} (b) provides
convincing evidence that
the long-time average approaches the thermal equilibrium value
from above like $1/L$, in agreement with {Eq.}~(\ref{28}).
Moreover, Fig.~\ref{fig5} (b) also provides reasonably good
evidence that the time-averaged 
variance $\sigma^2$ from {Eq.}~(\ref{29}) scales as $1/L^2$
(actually, the evidence is that the dispersion $\sigma$ scales as $1/L$),
as heuristically predicted in the previous paragraph,
and of course also 
in agreement with {Eqs.}~(\ref{32}), (\ref{35}).

Note that {the right-hand side of Eq.}~(\ref{15}) 
is a quasi-periodic function of $t$, meaning that there must 
exist time-points $t$ at which the expectation value 
returns arbitrarily close to its initial value (quantum recurrences).
Likewise, there must exist even more frequent temporal
``fluctuations'' which are smaller than those corresponding to such recurrences,
but which are still non-negligible (experimentally observable).
Only those are actually visible in Fig.~\ref{fig5} (b), since ``true'' recurrences
are way too rare to be observable on this time scale
(for the much smaller system in Fig.~\ref{fig2},
some of the fluctuations already come close 
to ``true'' recurrences).
Eqs.~(\ref{32}), (\ref{35}) predict, and Fig.~\ref{fig5} (b) numerically
illustrates that all the non-negligible deviations from the
thermal expectation value become more and more rare
as the system size increases, and provided one disregards
the initial relaxation phase. 
(The latter is in some sense quite similar to one of 
the above-mentioned recurrences.)
It is well-known that qualitatively, all these features also arise
in more general many-body systems via Eq.~(\ref{3}).
But to the best of our knowledge, 
our present theory (\ref{15}) captures 
them for the first time analytically in full quantitative detail.

A further interesting feature of Fig.~\ref{fig5} (b) is
the fact that the above-mentioned temporal
fluctuations are practically invisible
up to times $t$ of the order of $L$
(in our present units).
Put differently, the detailed $t$-dependence
for $t\leq\ord(L)$ remains (practically) unchanged 
upon further increasing $L$
(even in the limit $L\to\infty$), 
while the ``fluctuations'' 
for $t\geq\ord(L)$ still exhibit quite significant
pseudo-random changes upon increasing $L$.
A similar behavior has been often seen before in numerical 
studies, while our analytical result (\ref{15}) represents the
first quantitative explanation we know of.
Qualitatively, the usual heuristic argument
is \cite{lie72,sto95,bra06,ess16,duv19}
that up to times
until which the ``signal'' of our quantum quench at 
site $\nu$ (see Eqs.~(\ref{8}) and (\ref{14})) 
had sufficient time to reach both chain ends,
the system cannot ``know'' that the
distance between the ends is finite,
hence it behaves (nearly) as if the 
ends were infinitely far apart.
Assuming that the signal speed can be estimated by 
the pertinent Lieb-Robinson bound, 
it follows that the time in question
increases linearly with the system size.
Once again, this general 
heuristic argument
is nicely reproduced 
by our analytical finding (\ref{15}) according to 
Fig.~\ref{fig5} (b).
For additional analytical insight we also refer to
Sec.~\ref{s431}.

Yet another, even more subtle feature which has often 
been observed in previous numerical investigations
can also be seen in our present 
example in Fig.~\ref{fig5} (b):
Near $t=L$, where the almost $L$-independent behavior
goes over into the strongly $L$-dependent ``fluctuations'',
each curve seems to {exhibit} a minimum 
(on a slightly coarse-grained scale, or after 
taking some suitable running-time average).
In other words, the system transiently appears to be 
closer to thermal equilibrium than at even later times.
In hindsight, the same phenomenon can actually be 
observed already in Figs.~\ref{fig1}, \ref{fig3}, and \ref{fig4}:
Up to $t\simeq L=16$, the oscillations appear to steadily
die out, but later on they again become notably stronger.
While this apparently very common phenomenon
must be buried in Eq.~(\ref{15}), 
we did not succeed in pinning it down by purely 
analytical means.

\subsection{Non-thermalization after a global quench}
\label{s42}

As mentioned in Sec.~\ref{s2}, it is commonly expected 
{\cite{gog16,dal16,mor18,ued20,bar70}}
that our present (integrable) 
{model from Eq.}~(\ref{6}) generically does not exhibit thermalization after a 
global quench.
A particular example of this kind arises if the perturbation operator $V$
in {Eq.}~(\ref{8}) is chosen as
\begin{eqnarray}
V= \sum_{\nu=1}^L s_\nu^z
\ .
\label{36}
\end{eqnarray}
Though the 
{approximation} 
(\ref{13}) has not {yet} been rigorously 
{justified}
in such a case {(see also the discussion below Eq.~(\ref{13a}))}, 
it seems reasonable to conjecture that
the effects of weak perturbations are additive.
If so, we simply have to add up the effects of the single
$s_\nu^z$'s in {Eq.}~(\ref{14}) to obtain the pertinent results
for our present perturbation in {Eq.}~(\ref{36}).
Upon summing over all $\nu=1,...,L$ in {Eq.}~(\ref{15}) and
exploiting {Eqs.}~(\ref{18}), (\ref{20}) one thus obtains
\begin{eqnarray}
\langle s_\alpha^z \rangle_t - \langle s_\alpha^z\rangle _{\rm th}
=
g  \beta \sum_{n=1}^L (\tilde S_n^\alpha)^2  f_{nn}
\label{37}
\end{eqnarray}
independent of $t$.
Focusing on cases with $|\beta J|\ll 1$, 
the quantities $ f_{nn}$ can furthermore
be approximated by $1/[4\cosh^2(\beta \hh/2)]$,
as shown in Appendix \ref{appD} (see Eq.~(\ref{d8}) therein).
Together with {Eqs.}~(\ref{20}) and (\ref{37}) this yields
\begin{eqnarray}
\langle s_\alpha^z \rangle_t - \langle s_\alpha^z\rangle _{\rm th}
=
\frac{g \beta}{4\,\cosh^2(\beta \hh/2)}
\ .
\label{38}
\end{eqnarray}
Similarly as in
the previous subsection we thus can conclude
that the system indeed does {\em not} exhibit 
thermalization.

The specific perturbation $V$ from {Eq.}~(\ref{36}) has the 
advantage of being easily tractable and the shortcoming
that {Eq.}~(\ref{37}) is time-independent.
For more general perturbations which consist
of an extensive sum of single spins 
$s_\nu^z$, it is reasonable to expect a generally
non-trivial time-dependence of
$\langle s_\alpha^z \rangle_t - \langle s_\alpha^z\rangle _{\rm th}$.
By similar arguments as below Eqs.~(\ref{27}) and (\ref{36}),
one moreover expects that 
$\overline{\langle s_\alpha^z \rangle_t} - \langle s_\alpha^z\rangle _{\rm th}$
still approaches a non-vanishing value for large $L$,
which is again sufficient to rule out thermalization.

We reiterate that we employed an unproven (albeit reasonable) conjecture 
below Eq.~(\ref{36}).

As mentioned below Eq.~(\ref{11}), one 
might furthermore 
object that thermalization may possibly  still occur if
two different values of $\beta$ in {Eqs.}~(\ref{4}) and (\ref{7}) 
are chosen (but the two values still must be 
independent of the considered observable).

In conclusion, our above line of reasoning is far from a
rigorous demonstration of non-thermalization after a 
global quench, but it shows that our present
approach is at least 
not in contradiction to this widespread expectation
{\cite{gog16,dal16,mor18,ued20,bar70}}.

\subsection{Approximation for long chains}
\label{s43}

We begin by explaining the main ideas for a particularly 
simple special case.
Namely, we assume that
\begin{eqnarray}
|\beta J|\ll 1
\ .
\label{39}
\end{eqnarray}
As detailed in Appendix \ref{appD} (see Eq.~(\ref{d8}) therein), we thus can employ the approximation
\begin{eqnarray}
 f_{mn}\simeq  \frac{1}{4\cosh^2(\beta \hh /2)}
 =: \eta 
\label{40}
\end{eqnarray}
for all $m,n$.
Hence, {Eq.}~(\ref{15}) can be rewritten in the much simpler form
\begin{eqnarray}
\langle s_\alpha^z \rangle_t - \langle s_\alpha^z\rangle _{\rm th}
& = & 
g \beta \eta \, |W(t)|^2
\ ,
\label{41}
\\
W(t)
& := &
\sum_{n=1}^L \tilde S_n^\nu \tilde S_n^\alpha e^{{-i \tilde E_nt}}
\ .
\label{42}
\end{eqnarray}

Next we recall the symmetry under a sign change of $J$
and/or of $\hh$ as established 
below Eq.~(\ref{22}).
{Therefore}, we can and will henceforth replace 
$J$ by $|J|$ and $\hh$ by $| \hh |$.
Exploiting {Eqs.}~(\ref{16}), (\ref{19}), and the identity
\begin{eqnarray}
\sin(x)\sin(y)=[\cos(x-y)-\cos(x+y)]/2
\ ,
\label{43}
\end{eqnarray}
it 
{then}
follows from {Eq.}~(\ref{42}) that
\begin{eqnarray}
W(t) 
& = & 
e^{{i | \hh |t}}
\left[ \tilde W_{|\nu-\alpha|}(\omega t)-\tilde W_{\nu+\alpha}(\omega t) \right]
\ ,
\label{44}
\end{eqnarray}
{where}
\begin{eqnarray}
\omega
& := &
{|J|}
\label{45}
\end{eqnarray}
and
\begin{eqnarray}
\tilde W_k(t)
& := &
\frac{1}{L+1}\sum_{n=1}^L 
\cos(\mbox{$ \frac{\pi k n}{L+1}$}) 
\exp\{i t \cos(\mbox{$ \frac{\pi n}{L+1}$})\}
\ .
\label{46}
\end{eqnarray}
For large values of $L$ 
one thus obtains the asymptotic approximation
 \begin{eqnarray}
\tilde W_k  (t) 
& = & \frac{1}{\pi} \int_0^\pi dx  \cos(k  x)
\exp\{i t \cos(x)\}
\ .
\label{47}
\end{eqnarray}
The right-hand side can be readily related to
one of {the} various integral representations 
of the Bessel functions of the first kind $J_k(t)$
(see for instance Eq.~(9.1.21) in \cite{abram}), 
yielding 
\begin{eqnarray}
\tilde W_k (t) = i^k  J_k (t)
\ .
\label{48}
\end{eqnarray}

Our approximation of {Eq.}~(\ref{46}) by {Eq.}~(\ref{47}) is justified if the
summands in {Eq.}~(\ref{46}) change very little upon increasing $n$
by unity. The same requirement must apply to each of the 
two factors under the sum in {Eq.}~(\ref{46}), i.e., the two conditions
$k/L\ll 1$ and $t/L\ll 1$ must be satisfied.
In order to employ the approximation (\ref{47})
in {Eq.}~(\ref{44}) we therefore
must require that
\begin{eqnarray}
\omega t & \ll & L
\label{49}
\end{eqnarray}
and
\begin{eqnarray}
\nu+\alpha 
& \ll & L
\ .
\label{50}
\end{eqnarray}
Note that we also must require that $|\nu-\alpha|\ll L$, 
but this condition automatically follows
from {Eq.}~(\ref{50}) and the fact that
$\nu,\alpha\geq 1$ in {Eq.}~(\ref{14}).
Likewise, our assumption $L\gg 1$ above  Eq.~(\ref{47})
is automatically guaranteed 
by {Eq.}~(\ref{50}).

Given that {Eqs.}~(\ref{39}), (\ref{49}), and (\ref{50}) are fulfilled, 
we finally can infer from {Eq.}~(\ref{41}), (\ref{44}), (\ref{48}),
and the general identity $|\nu-\alpha|=\nu+\alpha-2\min\{\nu,\alpha\}$
the approximation
\begin{eqnarray}
\langle s_\alpha^z \rangle_t - \langle s_\alpha^z\rangle _{\rm th}
& = & 
g \beta \eta \, \left[ J_{|\nu-\alpha|}(\omega t)-\tau_\nu^\alpha J_{\nu+\alpha}(\omega t)\right]^2
,
\ \ \ \
\label{51}
\\
\tau_\nu^\alpha
& := &
(-1)^{\min\{\nu,\alpha\}}
\ .
\label{52}
\end{eqnarray}
Moreover, it is reasonable to expect that
this approximation will become 
asymptotically exact for $L\to\infty$
(while keeping $t$, $\nu$, and $\alpha$ fixed).

More precisely speaking, for any given (large but finite) $L$,
our approximation (\ref{47}) and all our subsequent 
conclusions will only apply up to moderately large times 
such that {Eq.}~(\ref{49}) is still fulfilled.
Most notably, they no longer depend explicitly
on $L$, only implicitly via the validity condition (\ref{49}).
For even larger times than in {Eq.}~(\ref{49}), 
we have to recourse to the original result 
in {Eq.}~(\ref{15})
(or in {Eq.}~(\ref{41})).
All this is closely related to and in agreement with the 
discussion in Sec.~\ref{s411}.

The asymptotic behavior of {Eq.}~(\ref{51}) for small and 
large times can be inferred from the well-known
properties of the Bessel functions $J_k(t)$,
which we collected for the sake of convenience in
Appendix \ref{appF}.
One thus obtains the approximations
\begin{eqnarray}
\langle s_\alpha^z \rangle_t - \langle s_\alpha^z\rangle _{\rm th}
& = &
g \beta \eta \, (\omega t/2)^{2|\nu-\alpha |}
\label{53}
\end{eqnarray}
for small $t$ and
\begin{eqnarray}
\langle s_\alpha^z \rangle_t - \langle s_\alpha^z\rangle _{\rm th}
& = &
\frac{8 g \beta \eta \nu^2\alpha^2}{\pi}\, \frac{\sin^2(\omega t\pm\pi/4)}{ (\omega t)^3}
\label{54}
\end{eqnarray}
for large $t$, where the plus signs applies if $\nu-\alpha$ is odd,
and the minus sign otherwise.
More precisely speaking, {Eq.}~(\ref{53}) requires
$\omega t\ll1$ (see also {Eq.}~(\ref{f1})), while
{Eq.}~(\ref{54}) requires 
{Eq.}~(\ref{49}) (see also previous paragraph). 
Moreover,
$\omega t \gg |(d -1/4)( d - 9/4)|$ with $d := (\nu-\alpha)^2$ 
(see below Eq.~(\ref{f9}))
is known to be yet another sufficient (but possibly not necessary)
prerequisite for the approximation of {Eq.}~(\ref{51}) by {Eq.}~(\ref{54}).

In view of {Eq.}~(\ref{14}), the condition (\ref{50}) implies that
both the perturbation and the observable are located
close to the left chain-end. By symmetry, analogous
results can be readily obtained if they are 
located near the right chain-end.

Next we turn to cases where the perturbation and the 
observable are not located near one of the chain ends.
For instance, both may be located near the
middle of the chain.
More generally, let us assume instead of {Eq.}~(\ref{50}) that
\cite{foot2}
\begin{eqnarray}
|\nu-\alpha| + 1
& \ll & L
\ ,
\label{55}
\end{eqnarray}
and that $\nu+\alpha$ increases with $L$ in such a way
that $(\nu+\alpha)/L$ 
is neither close to $0$ nor to $2$.
It follows that the approximation of {Eq.}~(\ref{46}) by
{Eq.}~(\ref{47}) is only justified for the first
function $\tilde W_{|\nu-\alpha|}(\omega t)$
on the right-hand side of {Eq.}~(\ref{44}).
On the other hand, it is now reasonable
to expect that the second function 
$\tilde W_{\nu+\alpha}(\omega t)$
can be approximately 
neglected in view of the fact that the first 
cosine term on the right-hand side of {Eq.}~(\ref{46})
exhibits relatively fast oscillations as a function 
of the summation index $n$ compared to the much slower
variations of the second cosine term (see also {Eq.}~(\ref{49})).
Hence, the right-hand side of {Eq.}~(\ref{46}) approaches zero
for large $L$.
Exploiting {Eq.}~(\ref{48}), this yields the approximation
\begin{eqnarray}
\langle s_\alpha^z \rangle_t - \langle s_\alpha^z\rangle _{\rm th}
= 
g \beta \eta \, [J_{|\nu-\alpha|}(\omega t)]^2
\label{56}
\end{eqnarray}
under {the} conditions (\ref{39}), (\ref{49}), 
and the one in and below Eq.~(\ref{55}).
For small $t$ this implies the same approximation as in {Eq.}~(\ref{53}),
while for large $t$ one now obtains
\begin{eqnarray}
\langle s_\alpha^z \rangle_t - \langle s_\alpha^z\rangle _{\rm th}
& = &
\frac{2 g \beta \eta }{\pi}\, \frac{\cos^2(\omega t\pm\pi/4)}{ \omega t}
\ ,
\label{57}
\end{eqnarray}
where, as in {Eq.}~(\ref{54}), the plus signs applies 
if $\nu-\alpha$ is odd, and the minus sign 
otherwise.
Furthermore, analogous remarks as below Eq.~(\ref{54}) 
and  in the paragraph below Eq.~(\ref{52}) apply.
In particular, 
$\omega t \gg |(\nu-\alpha)^2 - 1/4|$
(see below Eq.~(\ref{f8})) is 
known to be yet another
sufficient (but possibly not necessary)
prerequisite for the 
approximation of {Eq.}~(\ref{56}) by
{Eq.}~(\ref{57}).

\begin{figure}
\hspace*{-0.8cm}
\includegraphics[scale=0.95]{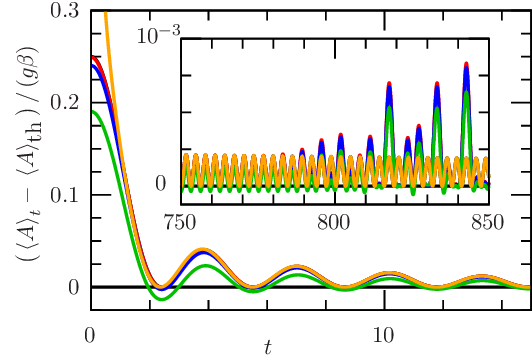}
\caption{
Numerical evaluation of Eq.~(\ref{15}) divided by 
$g\beta$ 
for $L=800$
and $\beta=0.2$ (red), 
$\beta=0.6$ (blue), 
$\beta=2$ (green).
The orange curve depicts the large-$t$ asymptotics from {Eq.}~(\ref{57}).
The approximation (\ref{56}) would be indistinguishable 
from the red curve in the main plot and from the 
orange curve in the inset, and is therefore not shown.
Everything else as in Fig.~\ref{fig1}.
}
\label{fig6}
\end{figure}

\begin{figure}
\hspace*{-0.8cm}
\includegraphics[scale=0.95]{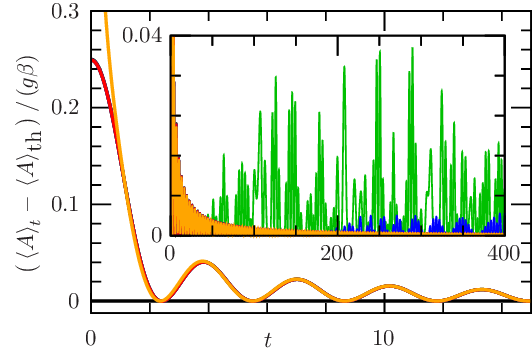}
\caption{
Same as in Fig.~\ref{fig6} but for $\beta=0.2$ and 
$L=800$ (red),  
$L=200$ (blue),  
$L=50$ (green).
Up to $t\approx 50$, the green curves are covered 
by the red curves, and likewise for the blue
curves up to $t\approx 200$.
In turn, the orange curves almost cover the red
curves between $t\approx 1$ and $t\approx 800$.
Beyond $t\approx 800$, the deviations between 
the red and orange curves would still be hardly 
discernible on the scale of the present inset, 
hence we actually plotted only $t$-values up 
to $400$.
}
\label{fig7}
\end{figure}

\begin{figure}
\hspace*{-0.8cm}
\includegraphics[scale=0.95]{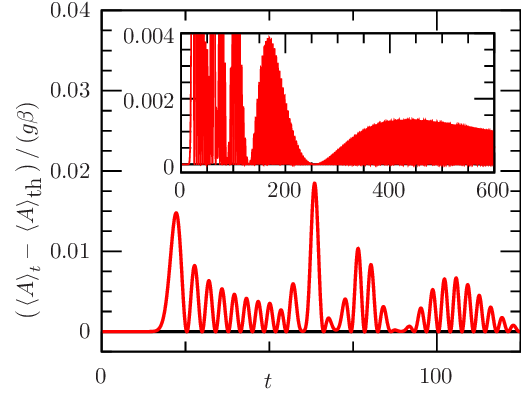}
\caption{
Numerical evaluation of Eq.~(\ref{15}) divided by 
$g\beta$
for $\nu=20$, $\alpha=40$, $L=800$, and $\beta=0.2$.
The approximation from {Eq.}~(\ref{51}) is indistinguishable 
from the red curve on the scale of this plot.
Everything else as in Fig.~\ref{fig1}.
}
\label{fig8}
\end{figure}

\subsubsection{Discussion}
\label{s431}

The prediction (\ref{56}) is mainly based on 
the three assumptions (\ref{39}), 
(\ref{49}), and (\ref{55}).
In practice, the condition (\ref{39}) essentially 
amounts to a small-$\beta$
(large temperature) approximation.
The main plot in Fig.~\ref{fig6} quantitatively illustrates this 
predicted convergence of {Eq.}~(\ref{15}) towards {Eq.}~(\ref{56})
upon decreasing $\beta$.
Furthermore, the approximation 
(\ref{57}) is depicted in orange.
Similarly, the main plot in Fig.~\ref{fig7} is meant to illustrate the
condition (\ref{55}), demonstrating the convergence 
of {Eq.}~(\ref{15}) towards {Eq.}~(\ref{56}) upon increasing $L$.
The breakdown of the approximation (\ref{57})
below $t\approx 1$ (orange curves 
in Figs.~\ref{fig6} and \ref{fig7}) 
can be understood 
via the prerequisite mentioned below Eq.~(\ref{57}).
On the other hand, the insets in Figs.~\ref{fig6} 
and \ref{fig7} numerically demonstrate that 
also the restriction to moderately large $t$
as quantified by {Eq.}~(\ref{49}) is indeed necessary
for the validity of the approximations (\ref{56}) and (\ref{57}).
For the rest, the findings in those insets are again closely related
to the discussion in the third paragraph of Sec.~\ref{s411}.

An analogous comparison of {Eq.}~(\ref{15}) with the
approximation (\ref{51}) is exemplified by Fig.~\ref{fig8}.
The main difference compared to all 
previous numerical examples is the choice 
of $\nu=20$ and $\alpha=40$, i.e., the
perturbation and the observable are neither close
to each other, nor close to the chain-end,
nor really deep in the ``bulk'' of the chain.
The first main message of Fig.~\ref{fig8} is
that the approximation (\ref{51}) agrees with
{Eq.}~(\ref{15}) very well at least for the depicted 
times $t \in[0,600]$.
On the other hand, the approximation (\ref{54})
is found to strongly disagree from the curves in 
Fig.~\ref{fig8} and is therefore not plotted.
Indeed, according to the extra prerequisite mentioned
below Eq.~(\ref{54}), a reasonably good agreement is
only expected for much larger $t$ values than those 
shown in Fig.~\ref{fig8}.
On the other hand, the good agreement of
{Eq.}~(\ref{15}) with {Eq.}~(\ref{51}) (and possibly with {Eq.}~(\ref{54}))
is expected to break down when {Eq.}~(\ref{49}) is 
violated, i.e., beyond $t\approx L$.
Since all this is similar as in Figs.~\ref{fig6} 
and \ref{fig7}, we did not show it once again 
in Fig.~\ref{fig8}.

Much more interesting appears to us to gain
a better understanding of the quite intricate
findings for $t \in[0,600]$ in Fig.~\ref{fig8}.
This is possible with the help of the following 
result from Appendix \ref{appF}:

For moderate-to-large values of $k$,
the Bessel functions $J_k(t)$ are negligibly small
for all $t< c \, k$ with some $k$-independent constant
$c$, which is numerically found to be close to unity
(and can be analytically lower bounded by 0.38, 
see Appendix \ref{appF}).
Similarly as at the end of
Sec.~\ref{s411}, this may be viewed as a consequence
and quantitative illustration of some suitable Lieb-Robinson bounds 
\cite{lie72,sto95,bra06,ess16,duv19}, 
where $1/c\approx 1$ plays
the role of a
speed at which the information that
there has been a quantum quench at time $t=0$ 
travels through the chain. 
Accordingly, both Bessel functions in {Eq.}~(\ref{51})
are negligible (practically no signal is observable)
up to times of the order of the distance $|\nu-\alpha|$ 
between perturbation and observable.
Subsequently, the first Bessel function in {Eq.}~(\ref{51})
is no longer negligible, while the second still remains
negligible up to times of the order $\nu+\alpha$.
The latter can be interpreted as the signal's traveling 
time from the perturbation site $\nu$ to the left
chain end, plus the traveling time of the reflected
signal to the measurement site $\alpha$.

At even later times, both Bessel functions
in {Eq.}~(\ref{51}) notably contribute and ultimately
result in the asymptotics (\ref{54}).
However, before this asymptotics actually sets in, 
quite complicated ``interference effects'' of 
the two Bessel functions in {Eq.}~(\ref{51}) are 
expected and indeed observed in Fig.~\ref{fig8}.

The only relevant remaining case is when 
$|\nu-\alpha|$ is large (implying that also
$\nu+\alpha$ and $2L-(\nu+\alpha)$ must be large).
Similarly as above Eq.~(\ref{56}) we thus may
consider both summands on the right-hand side
of {Eq.}~(\ref{44}) as negligible, in agreement with what 
one intuitively would have expected:
A local perturbation does not notably affect
sufficiently remote observables.

From the well-known properties of $J_k(t)$ collected in
Appendix \ref{appF} it seems reasonable to expect 
that $J_k(t)\to 0$ for $k\to\infty$ uniformly in $t$
(though we were not able to rigorously proof it).
If so, we see that (\ref{51}) already includes
(\ref{56}) as a special case.
In fact, (\ref{51}) will even remain valid
for arbitrary $\nu$ and $\alpha$, except
in situations where both $\nu$ and $\alpha$ 
are close to $L$.
(The behavior in the latter 
situations can be readily recovered 
by the symmetry argument in  the paragraph 
below  Eq.~(\ref{54}).)
The only remaining validity conditions for
{Eq.}~(\ref{51}) are then {Eq.}~(\ref{49}) and $L\gg 1$.
However, the different asymptotic results
in {Eqs.}~(\ref{54}) and (\ref{57}) indicate that things
are actually more subtle, at least
as far as the sequence of the
limits $\nu+\alpha\to\infty$ and $t\to\infty$ 
are concerned \cite{its93}.
Further such caveats are:
(i) The dependence on $|\nu-\alpha|$ in {Eq.}~(\ref{56})
has apparently completely disappeared from the
large-time asymptotics (\ref{57}).
However, this dependence still appears via the 
pertinent validity condition mentioned below  Eq.~(\ref{57}).
(ii) The asymptotics (\ref{54}) should go over into {Eq.}~(\ref{57}) upon 
increasing $\nu$ and $\alpha$ (provided {Eq.}~(\ref{55}) stays satisfied).
However, in doing so the factors $\nu^2\alpha^2$ in {Eq.}~(\ref{54}) 
increase quite strongly. 
On the other hand, {Eq.}~(\ref{54}) decreases as 
$t^{-3}$ and {Eq.}~(\ref{57}) only as $t^{-1}$.
Closer inspection shows that the validity conditions
below  Eqs.~ (\ref{54}) and (\ref{57}) are such that
the factors $\nu^2\alpha^2$ may indeed ``compensate'' 
the different powers of $t$ in just the right way to be 
still compatible with  a smooth transition between the 
two asymptotic results.

\subsection{Generalization}
\label{s44}

Throughout the previous Subsection \ref{s43} we assumed that the
condition (\ref{39}) is fulfilled.
Here, we briefly outline how one may 
proceed without this assumption.

The key idea is to consider instead of the original quantity 
in {Eq.}~(\ref{15}) its (negative) derivative
\begin{eqnarray}
v(t) := - \frac{d}{dt}(\langle s_\alpha^z \rangle_t - \langle s_\alpha^z\rangle _{\rm th})
\ .
\label{58}
\end{eqnarray}
Together with {Eqs.}~(\ref{17}), (\ref{18}), and (\ref{42}) we thus
can conclude (after a short calcuation) that
\begin{eqnarray}
v(t)
&=&
{g\,} 
\Im( W(t)  B^\ast(t) )
\ ,
\label{59}
\\
B(t)
& :=  &
\sum_{n=1}^L \tilde S_n^\nu \tilde S_n^\alpha \tanh(\beta \tilde E_n/2)\,
e^{{-i \tilde E_nt}}
\ .
\label{60}
\end{eqnarray}
Finally, we can integrate $v(t)$ from {Eq.}~(\ref{58}), yielding
\begin{eqnarray}
\langle s_\alpha^z \rangle_t - \langle s_\alpha^z\rangle _{\rm th}
=
\langle s_\alpha^z \rangle_{t'} - \langle s_\alpha^z\rangle _{\rm th}
+
\int_t^{t'} d\tau\, v(\tau)
\ .
\label{61}
\end{eqnarray}
According to the thermalization property from Sec.~\ref{s41}, 
the expectation values $\langle s_\alpha^z \rangle_{t'}$ must stay very 
close to $\langle s_\alpha^z\rangle _{\rm th}$
for the vast majority of all sufficiently 
late times $t'$ provided the system size 
$L$ is sufficiently large.
Hence the first two terms on the right-hand 
side of {Eq.}~(\ref{61}) can be considered to approximately 
cancel each other,
symbolically indicated as
\begin{eqnarray}
\langle s_\alpha^z \rangle_t - \langle s_\alpha^z\rangle _{\rm th}
=
\int_t^\infty d\tau\, v(\tau)
\ .
\label{62}
\end{eqnarray}

For large $L$, the function $W(t)$ in {Eq.}~(\ref{59}) can again be approximated
by {Eqs.}~(\ref{44}) and (\ref{47}) or (\ref{48}).
Likewise, the function $B(t)$ in {Eq.}~(\ref{60}) can be rewritten as
\begin{eqnarray}
B(t) 
& = & 
e^{{i | \hh | t}}
\left[ \tilde B_{|\nu-\alpha|}(\omega t)-\tilde B_{\nu+\alpha}(\omega t) \right]
\ ,
\label{63}
\end{eqnarray}
where $\tilde B_k(t)$ can be approximated for large $L$ as
\begin{eqnarray}
\tilde B_k(t)
& = &
\frac{1}{\pi}
\int_0^\pi dx  \cos(k x)\,
T_\beta(x)
\exp\{i t \cos(x)\}
\ ,
\label{64}
\\
T_\beta(x)
& := &
- \tanh(\beta[|J|\cos(x)+ |\hh| ]/2)
\ .
\label{65}
\end{eqnarray}

A central point of our present approach is that 
all three approximations
in {Eqs.}~(\ref{47}), (\ref{62}), and (\ref{64}) become 
asymptotically exact as $L\to\infty$.

In the absence of $T_\beta(x)$ in {Eq.}~(\ref{64}), 
the functions $\tilde B_k(t)$ would be identical to 
$\tilde W_k(t)$ from {Eq.}~(\ref{47}), 
and could be related to the Bessel 
functions $J_k(t)$ as in {Eq.}~(\ref{48}).
In the presence of $T_\beta(x)$ in {Eq.}~(\ref{64}),
an analogous relation to some well-established 
special functions does not seem to exist.
Put differently, we could consider {Eq.}~(\ref{64}) as the 
definition of a new special function and then determine 
its basic properties analogously to the basic properties of 
the Bessel functions in Appendix \ref{appF}.
However, this goes beyond the scope of our 
present paper, with the exception of one
particularly simple example, to which we now turn.

\begin{figure}
\hspace*{-0.8cm}
\includegraphics[scale=0.95]{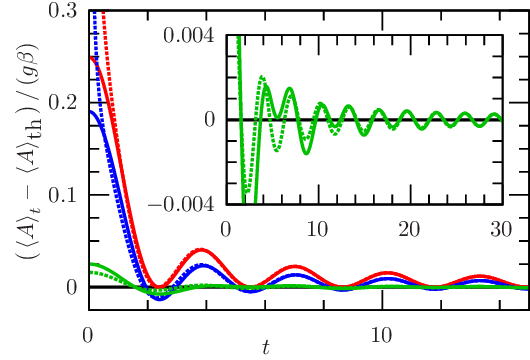}
\caption{
Solid:
Numerical evaluation of Eq.~(\ref{15}) divided by 
$g\beta$
for $L=800$
and $\beta=0.2$ (red), 
$\beta=2$ (blue), 
$\beta=20$ (green).
Everything else as in Fig.~\ref{fig1}.
Dotted:
The corresponding large-$t$
asymptotics from {Eq.}~(\ref{72}).
}
\label{fig9}
\end{figure}

As in {Eqs.}~(\ref{54}) and (\ref{57}) we focus on 
large times. Moreover, we take for granted 
the condition (\ref{49}), the conditions 
in and below Eq.~(\ref{55}) together with
\begin{eqnarray}
\alpha=\nu
\label{66}
\end{eqnarray}
(perturbation and observable identical and 
far from the chain ends),  and
the condition
\begin{eqnarray}
\hh =0
\label{67}
\end{eqnarray}
(no transversal field in {Eq.}~(\ref{6})).
Similarly as below Eq.~(\ref{55}), we thus can neglect the
last term in {Eq.}~(\ref{44}) and likewise for the last 
term in {Eq.}~(\ref{63}).
Utilizing {Eqs.}~(\ref{48}), (\ref{59}), and the fact that
the $J_k(t)$ are real functions of $t$ this yields
\begin{eqnarray}
v(t)= - 
{g\,}
J_0(\omega t)\Im( \tilde B_0(\omega t) )
\ .
\label{68}
\end{eqnarray}
As detailed in Appendix \ref{appG}, the explicit evaluation
of the right-hand side under the above specified conditions
is possible but somewhat tedious, resulting in the approximation
\begin{eqnarray}
v( t)
& = &
{\frac{2g}{\pi}}
\left(- \frac{b_0 c(t) s(t)}{\omega t} + \frac{b_1 c^2(t) - \frac{b_0}{8} s^2(t)}{(\omega t)^2}\right)
\ ,
\label{69}
\\
c(t) & := & \cos(\omega t-\pi/4)\, ,\ \ s(t) := \sin(\omega t-\pi/4) 
\, ,
\label{70}
\\
b_0 & := & -\tanh(\beta |J|/2)
\, , \ \ 
b_1=\frac{b_0}{8}+\frac{\beta |J|}{4\cosh^2(\beta J/2)}
\, ,
\ \ \ \ \ \ \ 
\label{71}
\end{eqnarray}
where corrections of the order of $(\omega t)^{-5/2}$
have been omitted on the right-hand side of {Eq.}~(\ref{69}).
Hence, the integral in {Eq.}~(\ref{62}) can be evaluated, yielding
\begin{eqnarray}
\langle s_\alpha^z \rangle_t - \langle s_\alpha^z\rangle _{\rm th}
& = & 
\frac{g}{\pi J}
\,
\frac{ \tanh(\frac{\beta J}{2})\, c^2(t) - r (\frac{\beta J}{2}) }{\omega t}
\label{72}
\end{eqnarray}
apart from higher order corrections in $1/\omega t$, and
where
\begin{eqnarray}
r(x) := \frac{\tanh(x)}{2}-
\frac{x}{2\cosh^2(x)}
\ .
\label{73}
\end{eqnarray}
One readily confirms that $r(x)$ is an odd and monotonically 
increasing function of $x$, scaling as $x^3$ for small $x$, and 
approaching $1/2$ for large $x$.
We also note that $|J|$ has been replaced by $J$ in {Eq.}~(\ref{72})
since the right-hand side is invariant under such a 
replacement.
Fig.~\ref{fig9} illustrates the quality of the 
large-$t$ asymptotics (\ref{72}).

In particular, for high temperatures (i.e. $|\beta J| \ll 1$, see also {Eqs.}~(\ref{5}) and (\ref{39}))
one recovers {Eq.}~(\ref{57}) from {Eq.}~(\ref{72}) (under the extra conditions (\ref{66}) and (\ref{67})).
Turning to low temperatures, one can deduce from {Eqs.}~(\ref{72}) and (\ref{73})
that
\begin{eqnarray}
\langle s_\alpha^z \rangle_t - \langle s_\alpha^z\rangle _{\rm th}
& = & 
\frac{g}{\pi J} 
\, 
\frac{\sin(2\omega t)}{\omega t}
\ \ \mbox{for $\beta J\gg 1$,}
\label{74}
\end{eqnarray}
and analogously for $\beta J \ll -1$.

More generally, the main difference between the high-temperature
approximation (\ref{41}) (see also {Eqs.}~(\ref{51})-(\ref{57}))
and the results (\ref{72}), (\ref{74}) for general temperatures
is that the former is strictly non-negative, while the later
exhibits sign changes,
see also Fig.~\ref{fig9}.

\section{Further examples}
\label{s5}

Generalizing {Eq.}~(\ref{14}), we now turn to
perturbations $V$ {and} observables $A$ 
of the form
\begin{eqnarray}
V=s_\nu^a\, , \ \ A=s_\alpha^b
\label{75}
\end{eqnarray}
with $a,b \in\{x,y,z\}$.

The case $a=b=z$ has been treated in Sec.~\ref{s4}.
In the case $a=z$ and $b\not=z$ one readily can infer
from {Eqs.}~(\ref{13}), (\ref{b43}), and (\ref{b44}) the trivial result
$\langle A\rangle_t = \langle A\rangle _{\rm th}=0$
independent of $t$.
In other words,  the perturbation does
not drive the observable out of equilibrium.
Similarly, one finds for $b=z$ and $a\not=z$ that
$\langle A\rangle_t = \langle A\rangle _{\rm th}$.
For a more detailed discussion of the latter 
expectation value
$\langle A\rangle _{\rm th}= \langle s^z_\alpha \rangle _{\rm th}$
we refer to Appendix \ref{appE}.

Altogether we thus can conclude that for any given 
perturbation
$V$ of the form $s_\nu^z$ the system exhibits 
thermalization at least as far as all single-site 
observables $A$ of the form {Eq.}~(\ref{75}) are 
concerned (see also Sec.~\ref{s2}).
The same applies to perturbations
of the form $V=s_\nu^{x,y}$ in combination with
observables of the form $A=s_\alpha^z$.

\begin{figure}
\hspace*{-0.8cm}
\includegraphics[scale=0.95]{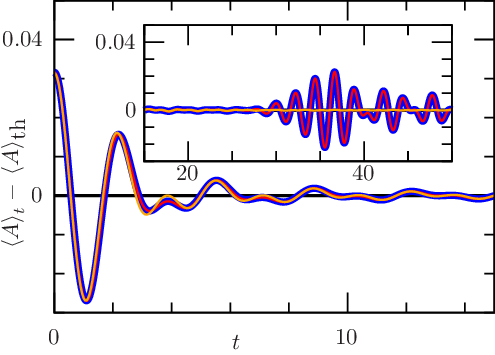}
\caption{
Blue: Numerically exact results for the same
system as in Fig.~\ref{fig1} (a) but now 
for $A=V=s_1^x$ and $\hh =2\sqrt{2}$.
Red: Analytical approximation from {Eq.}~(\ref{76}).
Orange: Approximation (\ref{81}).
Main plot: Initial relaxation behavior.
Inset: Transition from moderate- to long-time behavior.
}
\label{fig10}
\end{figure}

The remaining cases are 
(i) $a=b=x$ and (ii) $a=y$, $b=x$.
[The two other cases $a=b=y$ and $a=x$, $b=y$ 
then readily follow by symmetry arguments.]
In the special case $\nu=\alpha=1$
one obtains, similarly as in {Eq.}~(\ref{15}),
by means of the methods from Appendix 
\ref{appC} the result
\begin{eqnarray}
\langle s_1^x \rangle_t 
\! & - & \! 
\langle s_1^x\rangle _{\rm th}
= 
\frac{g \beta}{2}\Re (X (t))
\ \ \mbox{if $V=s_1^x$,}
\label{76}
\\
\langle s_1^x \rangle_t
\! & - & \! 
\langle s_1^x\rangle _{\rm th}
= 
\frac{g \beta}{2}\Im (X(t))
\ \ \mbox{if $V=s_1^y$,}
\label{77}
\\
X (t)
& := &
\sum_{n=1}^L (\tilde S_n^1)^2
\,
\frac{\tanh(\beta \tilde E_n/2)}{\beta \tilde E_n} 
\, 
e^{{-i \tilde E_nt}}
\ .
\ \ \ \ \
\label{78}
\end{eqnarray}
In the same way as below Eq.~(\ref{22}),
one can conclude that {Eq.}~(\ref{76})
must be an odd function of $\beta$,
and an even function of $J$ and of $\hh$,
while {Eq.}~(\ref{77}) must 
be an even function of $J$,
and an odd function of
$\beta$ and of $\hh$.
Fig.~\ref{fig10} exemplifies the very good 
agreement between the analytical 
approximation from {Eq.}~(\ref{76}) (red)
and numerically exact results (blue).

Similarly as in Sec.~\ref{s41} one can show that the long-time average
$\overline{X (t)}$ is a real number between zero and $1/L$, and that
$\overline{|X (t)|^2}=\overline{\Re(X ^2(t))}+\overline{\Im(X ^2(t))}$
is upper bounded by $1/2L$, implying that the observables in {Eqs.}~(\ref{76})
and (\ref{77}) exhibit thermalization.

Similarly as in Sec.~ \ref{s43}, taking for granted that {Eq.}~(\ref{39}) is
fulfilled we may employ the approximation
\begin{eqnarray}
\frac{\tanh(\beta \tilde E_n/2)}{\beta \tilde E_n} 
\simeq 
 \frac{\tanh(\beta \hh/2)}{\beta \hh} 
 = :\kappa
\ .
\label{79}
\end{eqnarray}
As in {Eqs.}~(\ref{41})-(\ref{46}) it then follows that
\begin{eqnarray}
X (t)
& = &
\kappa 
e^{{i \hh t}} 
\left[\tilde W_0 (\omega t) - \tilde W_{2}(\omega t)\right]
\label{80}
\end{eqnarray}
and that the functions $\tilde W_k(t)$ may be approximated by 
Bessel functions of the 
{first}
kind according to {Eq.}~(\ref{48}).
Note that, 
unlike in Eq.~(\ref{44}), the quantity $\hh$ must appear 
in Eq.~(\ref{80}) without the absolute sign to properly 
reproduce the symmetry properties below  
Eq.~(\ref{78}), see also Eq.~(\ref{45}).
Under the conditions (\ref{49}) and (\ref{50}) we thus obtain,
similarly as in {Eq.}~(\ref{51}), the approximation
\begin{eqnarray}
\langle s_1^x \rangle_t - \langle s_1^x\rangle _{\rm th}
& = & 
\frac{g\beta\kappa}{2}\, L(t)
\left[J_0 (\omega t)  +  J_{2}(\omega t)\right]
,
\ \ \ \ 
\label{81}
\end{eqnarray}
where
\begin{eqnarray}
L(t) := \left\{
  \begin{array}{l@{\hspace{3ex}}ll}
   \cos({\hh t}) & \mbox{if $V=s_1^x$} \\
    \sin({\hh t})  & \mbox{if $V=s_1^y$}
  \end{array}
  \right.
\label{82}
\end{eqnarray}
In the same way as in {Eq.}~(\ref{54}), this implies the large-$t$ asymptotics
\begin{eqnarray}
\langle s_1^x \rangle_t - \langle s_1^x\rangle _{\rm th}
& = & 
\sqrt{\frac{2}{\pi}}\,g\beta\kappa\,
L(t)
\frac{\sin(\omega t -\pi/2)}{(\omega t)^{3/2}}
\ .
\label{83}
\end{eqnarray}

The approximation (\ref{81}) is illustrated 
by the orange curve in Fig.~\ref{fig10}.
In accordance with the pertinent preconditions (\ref{39}), 
(\ref{49}), (\ref{50}), 
the agreement with the numerically exact
behavior (blue) is seen to be
very good up to times $t$ of the order of $L$.
[More precisely, the threshold is near $t=2L=32$.]
For even larger times, the situation is analogous
to the discussion at the end of Sec.~\ref{s411} 
and at the beginning of Sec.~\ref{s431}.
Similarly as in Fig.~\ref{fig6}, the approximation
(\ref{83}) would be indistinguishable from
{Eq.}~(\ref{81}) beyond $t=1$ in Fig.~\ref{fig10} and
is therefore not shown.
The specific choice of $G$ in Fig.~\ref{fig10}
results in oscillations of $L(t)$ in {Eq.}~(\ref{83})
with frequency 
${G}=2\sqrt{2}$
according to {Eq.}~(\ref{82}), while the
last term in {Eq.}~(\ref{83}) oscillates with
frequency $\omega=1$.
The resulting interference of the two
oscillations explains the quite non-trivial
time-dependence in the main plot of Fig.~\ref{fig10}.

Analogous results apply for the observable
$A=s_1^y$ instead of $A=s_1^x$,
and for the chain sites $\nu=\alpha=L$ instead 
of $\nu=\alpha=1$.
All the remaining cases are well-known to be
technically more demanding due to the non-locality
issues of the Jordan-Wigner transformation
pointed out at the end of Appendix \ref{appA1}.
Accordingly, explicit analytical
results for the correlations in {Eqs.}~(\ref{12}) and (\ref{13})
are only available in the literature for asymptotically
large or small values of $\beta$, 
see for instance in Refs.~\cite{cru81,sto92,sto95,der00} 
and further references therein.
Most notably, for $L\to\infty$, 
sufficiently small $\beta$, 
perturbations of the form $V=s_\nu^x$,
and under the conditions in and around  Eq.~(\ref{55}) 
one finds the asymptotic approximation
\begin{eqnarray}
\langle s_\alpha^x \rangle_t - \langle s_\alpha^x\rangle _{\rm th}
& = & 
\delta_{\nu \alpha} \frac{g\beta}{4}  
\cos({\hh t}) 
e^{-\omega^2 t^2/4}
\ .
\label{84}
\end{eqnarray}
Moreover, for sufficiently large times
this superexponential  decay is known to ultimately 
cross over into an ordinary exponential decay 
whenever $\beta\not = 0$ \cite{its93,sto95}.

\begin{figure}
\hspace*{-0.8cm}
\includegraphics[scale=0.95]{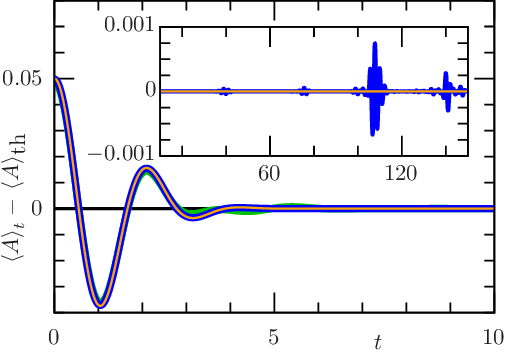}
\caption{
Blue: Numerically exact results for the same
system as in Fig.~\ref{fig1} (a) but now for 
$A=V=s_{L/2}^x$, $\hh =2\sqrt{2}$,
$g=4$, and $\beta=0.05$.
Green: same but for $g=1$ and $\beta=0.2$.
Orange: Large-$L$ and small-$\beta$ asymptotics from {Eq.}~(\ref{84}).
Main plot: Initial relaxation behavior.
Inset: Transition from moderate- to 
long-time behavior 
(the green curve is omitted).
}
\label{fig11}
\end{figure}

A comparison of the approximation (\ref{84}) 
with numerically exact results is provided in 
Fig.~\ref{fig11}.
Similarly as in Fig.~\ref{fig10}, the agreement 
is very good up to times $t$ of the order of $L$,
but not any more for substantially larger times (inset),
while the details of the latter disagreement are quite
different in the two figures.
Note that a theoretical prediction corresponding to the
red curve in Fig.~\ref{fig10} is not available
in the case of Fig.~\ref{fig11}.
Moreover, the approximation (\ref{84})
requires $\beta$ to be even smaller than 
the corresponding approximation (\ref{81})
in  Fig.~\ref{fig10}, as can be inferred upon comparison 
of the green and blue curves in Fig.~\ref{fig11}.

Since analytical results for finite $L$ are not available 
in our present case, we cannot verify thermalization
in the same way as in our previous examples.
However, if one takes for granted that the limits
$L\to\infty$ and $t\to\infty$ can be exchanged
(as it is often done),
and that the derivation of the above-mentioned 
results can be considered as sufficiently rigorous, 
thermalization follows immediately from those results.

We recall that {Eq.}~(\ref{84}) applies if $V=s_\nu^x$.
Upon comparison with {Eq.}~(\ref{81}) it is natural to conjecture
that a sine instead of the cosine will appear on the 
right-hand side of {Eq.}~(\ref{84}) if $V=s_\nu^y$.
Appendix \ref{appI} provides a more careful
justification of this conjecture (see Eq.~(\ref{i1}) therein).

\subsection{{Synopsis and extensions}}
\label{s51}

{Let us finally collect the various findings 
of this and the previous sections.
As stated below Eq.~(\ref{75}), 
we covered all single-site observables 
$A=s_\alpha^b$ for arbitrary perturbations 
of the form $V=s_\nu^z$.
In particular, thermalization and various more detailed
properties of the time-dependent expectation values
have been deduced under the proviso that the parameter
$\epsilon$ in Eq.~(\ref{13a}) remains sufficiently small.
With respect to the remaining cases mentioned
above Eq.~(\ref{76}), we also covered
many particularly interesting single-site 
observables for perturbations of the form 
$V=s_\nu^{x,y}$.
An analogous treatment of the remaining single-site 
observables is expected to be in principle straightforward,
but in practice the required calculations are,
as explained above Eq.~(\ref{84}), 
well known to be a very daunting task, which 
goes beyond the scope of our present work.
Essentially the same applies to
observables $A$ or perturbations $V$
which consist of a product of several 
single-spin operators.}

{Further modifications and extensions, 
which are expected to be technically less challenging 
but nevertheless omitted here, include 
XX-models with periodic 
instead of the open boundary conditions 
in Eq.~(\ref{6}), and so-called XY-models 
with different couplings of the nearest neighbor
interactions of the $x$- and $y$-spins in Eq.~(\ref{6}),
see also Refs~\cite{bar70,lie61}.}

{Finally, yet another kind of generalization is 
obvious, yet noteworthy.
Given some perturbation $V$ with its above
specified class of ``admitted'' observables $A$, 
arbitrary linear combinations of those observables 
will obviously be admitted as well.
Formally, the reason is that Eq.~(\ref{2}) is linear in $A$.
Moreover, also the quantitative properties
(including thermalization) immediately follow
by linearity from the known properties of the 
summands.
Likewise, given some observable $A$ with
its concomitant class of ``admitted'' perturbations $V$, 
linear combinations of those perturbations 
are again admissible due to the fact that the 
right-hand side of Eq.~(\ref{13}) is linear in $V$.
Moreover, the quantitative behavior of the given
observable $A$ again follows by linearity.
However, an additional requirement is now
that the parameter $\epsilon$ in Eq.~(\ref{13a}) 
must still be small for the perturbation $V$ under 
consideration.
In particular, extensive perturbations 
(corresponding to a global quench scenario)
are thus
excluded in the thermodynamic limit $L\to\infty$,
as already mentioned below 
Eq.~(\ref{13a}) and in Sec.~\ref{s42}.}

\section{Conclusions}
\label{s6}

This section complements the general considerations 
in Secs.~\ref{s1} and \ref{s2}, which are therefore not
repeated here.

The two main achievements of our present work are a rigorous
demonstration of thermalization and a detailed analytical
as well as numerical exploration of the temporal 
relaxation behavior.
For this purpose we focused on a particularly simple model,
namely the $XX$-spin chain from {Eq.}~(\ref{6}),
{on} particularly simple initial states,
essentially amounting to weak local 
quantum quenches,
{and on single-site observables and sums thereof, 
as detailed in the previous Sec.~\ref{s51}.}
On the one hand, one may say that these assumptions 
and restrictions are rather strong.
On the other hand, in view of how little is rigorously 
known with regard to thermalization (see also Sec.~\ref{s2}), 
it seems unavoidable to get used to some degree 
of modesty in this context.

Despite the simplicity of the model, its detailed
temporal behavior turned out to be very rich,
ranging from damped oscillations 
with a possibly very slow algebraic decay,
to complicated interferences of several such damped 
oscillations, to a superexponential (Gaussian)
decay in time.
The latter is ultimately caused by the non-local character 
of the pertinent Jordan-Wigner transformation 
and shows that our present simple example already
goes beyond the realm of ``effectively'' non-interacting
theories in combination with local observables,
see e.g. Refs.~\cite{mur19,glu19} and numerous further 
references cited therein.
Moreover, those non-interacting theories are also 
complementary to our present work in that they 
cover a considerably more general class of (integrable)
models 
and initial conditions, while no analytical results with
respect to the issue of thermalization are obtained
(instead, so-called generalized Gibbs ensembles
often play a central role).

A specific virtue of our present approach is 
that one can explore in quite some detail 
the system's behavior for a possibly very large but 
fixed time-point $t$ when {the} chain-length $L$
becomes arbitrarily large (thermodynamic limit),
but also
for a possibly very large but 
finite $L$ when $t$ becomes arbitrarily large.
Most importantly, the behavior in one case
turned out to hardly admit any conclusions 
regarding the other case
(see Secs.~\ref{s411}, \ref{s431}, and \ref{s5}).
Therefore, the relevance of analytical results 
in the thermodynamic limit
with respect to the long-time behavior of real (finite)
systems may be less obvious
than what is often suggested in the literature.

We furthermore remark that there 
exists a considerable variety
of related but different analytical investigations,
dealing with entire statistical ensembles of models
\cite{foot4}, and resulting in predictions which apply with very
high probability to a 
{\em randomly sampled} member 
of the ensemble.
However, and in contrast to our present approach,
with respect to a
{\em given} (non-random) model system, 
there always remains some uncertainty.
Even more serious issues are:
(i) The same, given model may
possibly be viewed as
a member of two different ensembles, 
each of which predicts 
a different behavior with high probability.
(ii) In order to be analytically tractable,
certain basic properties of the given model are often
no longer true for the vast majority of all the other 
members of the ensemble.
For instance, the system of actual interest 
may exhibit only short-range and few-body interactions
(such as our present model from {Eq.}~(\ref{6})),
but {\em not} the vast majority of the other members 
of the ensemble.

In our present study we 
{mainly focused on}
quenches 
and observables each of which acts nontrivially on only 
one site of the spin chain (and on sums of such observables).
Various generalizations are in principle straightforward, 
while the technical details (for instance in 
Appendix \ref{appB1}) will {often} be 
substantially more involved, {see also Sec.~\ref{s51}}.

Further extensions which may be worthwhile to be explored in
the future 
include generalizations beyond the realm of weak local
perturbations and of our simple XX-spin-chain model (\ref{6}).
Specifically, we are aware of a variety of non-rigorous arguments
according to which 
thermalization after a {\em local} quench might appear as being 
almost trivial even for much more general models than in {Eq.}~(\ref{6}).
However, such arguments 
may be misleading in view of the following result,
which will be elaborated in a separate work.
Namely, {\em absence} of 
thermalization after a local quench 
can be analytically shown to occur 
for a slight modification of our present 
setup in Eqs.~(\ref{6})-(\ref{8}).
In other words, minor details may be of great 
importance regarding the issue of thermalization,
and rigorous results are therefore quite valuable
even for apparently very simple models.

\begin{acknowledgments}
We thank Jannis Eckseler for valuable numerical assistance.
This work was supported by the 
Deutsche Forschungsgemeinschaft (DFG, German Research Foundation)
under Grant No. 355031190 
within the Research Unit FOR 2692
and under Grant No. 
502254252, 
and
by the Paderborn Center for Parallel 
Computing (PC$^2$) within the project 
HPC-PRF-UBI2.
\end{acknowledgments}

\vspace*{0.5cm}
\appendix

\section{Jordan-Wigner Transformation and beyond}
\label{appA}

This Appendix summarizes the 
main steps of the so-called Jordan-Wigner 
transformation,
and discusses some implications of
specific interest in our present paper.
Since this transformation is well-established basic knowledge, 
we only mention here once and for all a
very incomplete selection of pertinent previous works,
see Refs.~{\cite{cru81,sto92,sto95,der00,bar70,lie61}} and further
references therein.
On the other hand, we will include sufficient details to make the 
calculations self-contained.
Since such calculations may also be of interest in other 
contexts, we actually consider somewhat more general 
model Hamiltonians than in Eq.~(\ref{6}) of 
the main paper, namely
\begin{eqnarray}
H = 
- \sum_{l=1}^{L-1}  J_l \, (s^x_{l+1}s^x_l  + s^y_{l+1}s^y_l)
- \sum_{l=1}^{L}  \hh_l\, s_l^z
\, .
\label{a1}
\end{eqnarray}
Similarly as below  Eq.~(\ref{6}),
it is well-known (and can be readily verified 
by means of the subsequent formalism)
that $S^z:=\sum_{l=1}^L s_l^z$
is still a conserved quantity, i.e., $[H,S^z]=0$,
for arbitrary $J_l$ and $\hh_l$.

\subsection{Standard Jordan-Wigner Transformation}
\label{appA1}

As usual (see also below Eq.~(\ref{6})), it is convenient to employ the 
eigenbasis of the spin operators $s_l^z$,
implying that $s^{a}_l=\sigma_l^a/2$,
where the $\sigma_l^a$ are Pauli-matrices
and $a\in\{x,y,z\}$.
Moreover,
all operators are understood as acting on the system's
full many-body Hilbert space (tensor product of $L$ spin-1/2 spaces), hence 
Pauli matrices with different indices $l$ 
always commute  with each other.

Defining the usual raising and lowering operators as
$\sigma_l^{\pm}:=(\sigma_l^x\pm i\sigma_l^y)/2$
and exploiting that the $\sigma_l^a$ are Pauli matrices one readily verifies that
\begin{eqnarray}
\sigma_l^x & = & \sigma_l^+ + \sigma_l^-
\ ,
\label{a2}
\\
\sigma_l^y & = & (\sigma_l^+ - \sigma_l^-)/i
\ ,
\label{a3}
\\
\sigma_l^z & = &  
2\sigma_l^+\sigma_l^- -1
\ ,
\label{a4}
\\
\sigma_l^z \sigma_l^\pm & = & \pm \sigma_l^\pm 
\ .
\label{a5}
\end{eqnarray}
Hence, the Hamiltonian (\ref{a1}) can be rewritten as
\begin{eqnarray}
H  &=& \tilde H + \hh_{tot} 
\label{a6}
\end{eqnarray}
with
\begin{eqnarray}
\tilde H := -\frac{1}{2}\sum_{l=1}^{L-1} J_l
(\sigma_{l+1}^+ \sigma_l^- + \sigma_{l}^+ \sigma_{l+1}^-)
-\sum_{l=1}^L \hh_l \,
\sigma_l^+\sigma_l^-
\ \ \ \ \ \ 
\label{a7}
\end{eqnarray}
and
\begin{eqnarray}
\hh_{tot} := \frac{1}{2}\sum_{l=1}^L\hh_l
\ .
\label{a8}
\end{eqnarray}

Next we define the so-called string operators
\begin{eqnarray}
Z_l:=\prod_{k=1}^{l-1} (-\sigma_k^z)
\label{a9}
\end{eqnarray}
and the creation and annihilation operators
\begin{eqnarray}
c_l^\dagger & := & \sigma^+_l Z_{l}
\ ,
\label{a10}
\\
c_l & := & (c_l^\dagger)^\dagger = Z_{l}\, \sigma^-_l 
\ .
\label{a11}
\end{eqnarray}
Some useful technical side remarks are: 
(i) $Z_{1}=1$,
 implying $c_1^\dagger=\sigma_1^+$ and $c_1=\sigma_1^-$.
(ii) $Z_{l}$ commutes with $\sigma_l^{\pm}$, implying
$c_l^\dagger = Z_{l}\, \sigma^+_l$ and $c_l = \sigma^-_l Z_{l}$.
(iii) Since $(\sigma_l^z)^2=1$ we find $Z_l^2=1$, implying 
$\sigma_l^+=Z_{l}\,c_l^\dagger=c_l^\dagger Z_{l}$, 
$\sigma_l^-=Z_{l}\,c_l =c_l Z_{l}$,
and $\sigma_l^+\sigma_l^-=c_l^\dagger c_l$.
(iv) Exploiting {Eq.}~(\ref{a5}) it follows that
$\sigma_{l+1}^+\sigma_l^-=c_{l+1}^\dagger c_l$.

We thus can rewrite {Eqs.}~(\ref{a2})-(\ref{a4}), (\ref{a7}), and (\ref{a9})
as 
\begin{eqnarray}
\sigma_l^x & = & Z_{l}(c_l^\dagger+c_l)\ ,
\label{a12}
\\
\sigma_l^y & = & Z_{l}(c_l^\dagger-c_l)/i
\ ,
\label{a13}
\\
\sigma_l^z & = &  2c_l^\dagger c_l-1
\ ,
\label{a14}
\\
\tilde H & = & 
- \frac{1}{2}\sum_{l=1}^{L-1}  J_l \, (c_{l+1}^\dagger c_{l}+c_l^\dagger c_{l+1})
- \sum_{l=1}^{L}  \hh_l\, c_l^\dagger c_l 
, \ \ \ \ \ 
\label{a15}
\\
Z_l & = & \prod_{k=1}^{l-1}(1 - 2c_k^\dagger c_k)
\ .
\label{a16}
\end{eqnarray}

Similarly as in {Eqs.}~(\ref{a2})-(\ref{a5}) 
one readily verifies that
\begin{eqnarray}
\{\sigma_l^+,\sigma_l^-\}
& = &
1
\ ,
\label{a17}
\\
\{\sigma_l^+,\sigma_l^+\}
& = & \{\sigma_l^-,\sigma_l^-\} =
0
\ ,
\label{a18}
\end{eqnarray}
where 
$\{A,B\}:=AB+BA$ (anti-commutator).
Furthermore, by combining {Eq.}~(\ref{a5}) with its adjoint counterpart 
$\sigma_l^\mp \sigma_l^z= \pm \sigma_l^\mp$
one sees that
\begin{eqnarray}
\sigma_l^z   \sigma_l^\pm = -  \sigma_l^\pm \sigma_l^z 
\ ,
\label{a19}
\end{eqnarray}
and with {Eq.}~(\ref{a9}) that
\begin{eqnarray}
Z_k\sigma_l^\pm  =  \left\{
  \begin{array}{l@{\hspace{3ex}}ll}
      \sigma_l^\pm Z_k &\mbox{if $k\leq l$} \\
    - \sigma_l^\pm Z_k\ &\mbox{if $k>l$}
  \end{array}
  \right.
\label{a20}
\end{eqnarray}
Together with {Eqs.}~(\ref{a10}), (\ref{a11}), (\ref{a17}), (\ref{a18})
we thus can infer
the anti-commutation relations
\begin{eqnarray}
\{c_k^\dagger,c_l\}=\delta_{kl}\, ,\ 
\{c_k,c_l\}=\{c^\dagger_k,c^\dagger_l\}=0 
\label{a21}
\end{eqnarray}
for all $k,l\in\{1,...,L\}$,
where $\delta_{kl}$ is the Kronecker delta. 
In other words, the $c_l^\dagger$ and $c_l$ are 
indeed fermionic creation and annihilation operators.

We remark that the operators $\sigma_l^\pm$ 
fulfill analogous anti-commutation relations as in {Eq.}~(\ref{a21}) if $k=l$ 
(see {Eqs.}~(\ref{a17}) and (\ref{a18})),
but in general not any more if $k\not=l$.
(Instead, they fulfill the commutation relations
$[\sigma_k^+,\sigma_l^-]=\delta_{kl}\sigma_k^z$,
$[\sigma_k^+,\sigma_l^+]=0$, and
$[\sigma_k^-,\sigma_l^-]=0$.)
This is the main advantage of the above Jordan-Wigner
transformation from the operators
$\sigma_l^+$ and $\sigma_l^-$ to the operators $c_l^\dagger$ 
and $c_l$.
A notable disadvantage is that  
{Eqs.}~(\ref{a9})-(\ref{a11}) are, in general, non-local operators 
(with respect to the actual model of interest in {Eq.}~(\ref{a1})).
Analogously, the basic observables {Eqs.}~(\ref{a12}) and (\ref{a13})
of the original model are generally non-local operators
of the transformed model (see also {Eq.}~(\ref{a16})).

More precisely speaking, since $Z_1=1$ (see below  Eq.~(\ref{a11}))
the transformed observables {Eqs.}~(\ref{a12}) and (\ref{a13}) remain local
for $l=1$, and similarly for a few more (small) $l$-values
(near the left chain end).
Likewise, by replacing $\prod_{k=1}^{l-1}$ on the right-hand side of {Eq.}~(\ref{a9})
by $\prod_{k=l+1}^L$, the transformed observables 
would remain local near the right chain end 
(more physically, this essentially amounts to interchanging 
the chain ends). 
However, away from both chain ends, the non-locality
remains a non-trivial issue, see also the discussion 
above Eq.~(\ref{84}) in the main paper.

\subsection{Diagonalization}
\label{appA2}

Instead of the specific Hamiltonians from {Eq.}~(\ref{a15}) 
let us temporarily consider 
a general ``quadratic'' 
Hamiltonian of the form
\begin{eqnarray}
\tilde H = \sum_{k,l=1}^L B_{kl} \, c_k^\dagger c_l
\ ,
\label{a48}
\end{eqnarray}
with arbitrary coefficients $B_{kl}$ apart from the hermiticity
condition $B_{lk}=B_{kl}^\ast$.
Considering the $B_{kl}$ as the matrix elements of a 
Hermitian $L\times L$ matrix $B$, it follows that 
there exists a unitary matrix $U$ with the property
$U^\dagger D U = B$, where $D$ is a diagonal 
Hermitian matrix.
In other words its matrix elements are of the form
$D_{kl}=\delta_{kl} \tilde E_k$ with real diagonal elements 
$\tilde E_k$.
Denoting by $c$ a column vector (or vector-operator)
with entries (elements) $c_l$, and by $c^\dagger$ 
the concomitant row vector with
entries $c_l^\dagger$, we can rewrite {Eq.}~(\ref{a48})
as $\tilde H = c^\dagger B c =  f^\dagger D  f$,
where $ f:=Uc$ is a column vector with entries
\begin{eqnarray}
 f_k:=\sum_{l=1}^L U_{kl} c_l
\ , 
\label{a49}
\end{eqnarray}
and $ f^\dagger$ the concomitant row vector with entries 
$ f_k^\dagger=( f_k)^\dagger$. 
In other words, {Eq.}~(\ref{a48}) takes the form
\begin{eqnarray}
\tilde H = \sum_{k=1}^L \tilde E_k \,  f_k^\dagger  f_k
\ .
\label{a50}
\end{eqnarray}
Exploiting {Eqs.}~(\ref{a21}) 
and (\ref{a49})
we can conclude that
\begin{eqnarray}
\{ f_k^\dagger,  f_l\}=\delta_{kl}\, ,\ 
\{ f_k,  f_l\}=\{ f^\dagger_k,  f^\dagger_l\}=0 \ ,
\label{a51}
\end{eqnarray}
i.e., the $ f_l^\dagger$ and $ f_l$ 
are again fermionic creation and annihilation operators.

Let us denote by $|0\rangle:=|\!\downarrow \cdots\downarrow\rangle$
the specific $L$-spin product state where every single spin is 
in the ``down'' state 
in the eigenbasis of $s_l^z$, i.e.,
$\sigma_l^z |0\rangle=-|0\rangle$ for all $l$.
Exploiting that $\sigma_l^-:=(\sigma_l^x-i\sigma_l^y)/2$
(see above  Eq.~(\ref{a2})) one can infer that
$\sigma_l^-|0\rangle=0$
for any $l\in\{1,...,L\}$, implying with {Eq.}~(\ref{a11}) that
$c_l|0\rangle = 0$ and with {Eq.}~(\ref{a49}) that
\begin{eqnarray}
 f_l|0\rangle = 0
\label{a52}
\end{eqnarray}
for any $l\in\{1,...,L\}$.
Alternatively, $|0\rangle$ may thus be viewed 
as ``vacuum state'', especially with reference to fermionic 
Hubbard-like models as in {Eq.}~(\ref{a15}).

Next we denote by 
{$\vec b:=(b_1,...,b_L)$}
a vector 
with $L$ 
{``binary'' components $b_k\in\{0,1\}$,
i.e., there are $2^L$ different such vectors.
For any given $\vec b$},
we furthermore define the state 
\begin{eqnarray}
|\vec b\rangle := 
( f_1^\dagger)^{b_1}
( f_2^\dagger)^{b_2}
\cdots 
( f_L^\dagger)^{b_L}
\,  |0\rangle
\ .
\label{a53}
\end{eqnarray}
[Note that the order of the factors on the right-hand 
side is important since
they generally do not commute according to {Eq.}~(\ref{a51}).]
As usual, the occupation (or particle) number operator is defined
as
\begin{eqnarray}
n_k:= f_k^\dagger  f_k
\ .
\label{a54}
\end{eqnarray}
With {Eq.}~(\ref{a51}) it follows that 
$n_kf^\dagger_k=f^\dagger_k -f^\dagger_k n_k$
and $n_kf^\dagger_l=f^\dagger_ln_k$
if $k\not=l$.
Together with {Eqs.}~(\ref{a52}) and (\ref{a53}) 
we thus can conclude that 
\begin{eqnarray}
n_k|\vec b\rangle = b_k |\vec b\rangle
\label{a55}
\ ,
\end{eqnarray}
i.e., $n_k$ indeed 
counts the number of particles (fermions) in state $k$.
Moreover, we can infer from {Eq.}~(\ref{a50})
that
\begin{eqnarray}
\tilde H \,|\vec b\rangle & = & \tilde E(\vec b)\, |\vec b\rangle
\ ,
\label{a56}
\\
\tilde E(\vec b) & := & \sum_{k=1}^L b_k\, \tilde E_k
\ ,
\label{a57}
\end{eqnarray}
and that $\langle \vec b'|\vec b\rangle=\delta_{\vec b'\!\vec b}$,
where $\delta_{\vec b'\!\vec b}:=1$ if 
$\vec b'=\vec b$ and $\delta_{\vec b'\!\vec b}:=0$ otherwise 
(generalized Kronecker delta).

In other words, we have obtained an orthonormalized set of 
$2^L$ eigenstates of $\tilde H$, i.e., we have formally solved 
the eigenvalue problem for any Hamiltonian 
$\tilde H$ of the general form (\ref{a48}). 

Note that our present energies $\tilde E_k$ and matrix 
elements $U_{kl}$ have a different meaning than the 
energies and matrix elements appearing in Sec.~\ref{s2} 
of the main paper (see in and around Eq.~(\ref{3})):
The latter refer to properties of the full many-body 
Hamiltonian, while 
the former refer to certain single particle properties.
In particular, the many-body energies $E_n$ in Sec.~\ref{s2}
essentially correspond to the energies $\tilde E(\vec b)$ in 
{Eq.}~(\ref{a57}), albeit a different labeling of those energies 
by $n$ and by $\vec b$, respectively, is employed in the two cases.

{We finally note that any given energy 
eigenstate in Eq.~(\ref{a53}) (apart from $|0\rangle$) 
can be rewritten as
\begin{eqnarray}
|k_1,...,k_n\rangle :=  f_{k_1}^\dagger \cdots f_{k_n}^\dagger | 0 \rangle
\label{x2}
\end{eqnarray}
for some suitably chosen $n\in\{1,...,L\}$ and
$k_1,...,k_n\in\{1,...,L\}$ with
$k_1<k_2<...<k_n$.
In the context of our Hubbard-like models from Eq.~(\ref{a15})
with vacuum state $| 0 \rangle$ (see below Eq.~(\ref{a52})),
such an energy eigenstate (\ref{x2}) is
-- for obvious reasons -- often denoted as 
``$n$-particle excitation''.
Exploiting Eqs.~(\ref{a10}), (\ref{a20}), and (\ref{a49})
we thus can conclude that
\begin{eqnarray}
|k_1,...,k_n\rangle 
=  
\mbox{$\sum'$}
\tau({\bf l})\, 
U^\ast_{k_1l_1} \cdots U^\ast_{k_nl_n}
\sigma^+_{l_1} \cdots \sigma^+_{l_n} | 0 \rangle
\, ,
\ \ \ \ \ \ 
\label{x3}
\end{eqnarray}
where $\sum'$ indicates a sum over all pairwise different indices
$l_1,...,l_n\in\{1,...,L\}$, 
and where $\tau({\bf l})$ is either $+1$ or $-1$,
but for the rest depends in a very complicated way on 
${\bf l}:=(l_1,...,l_n)$ as a consequence of Eq.~(\ref{a20}).
Observing that all $\sigma^+_l$ commute with each other, this implies
\begin{eqnarray}
|k_1,...,k_n\rangle 
=  
\mbox{$\sum''$}
c(l_1,...,l_n
)\,
\sigma^+_{l_1} \cdots \sigma^+_{l_n} | 0 \rangle
\, ,
\ \ \ \ \ \ 
\label{x4}
\end{eqnarray}
where $\sum''$ indicates a sum over all  $l_1,...,l_n\in\{1,...,L\}$
with $l_1<l_2<...<l_n$ and thus consists of $\binom{L}{n}$ summands.
Furthermore,
\begin{eqnarray}
c(l_1,...,l_n) := \sum_\Pi 
\tau(\Pi({\bf l})) 
\, 
U^\ast_{k_1l_{\Pi(1)}} \cdots U^\ast_{k_nl_{\Pi(n)}}
\, , \ \ 
\label{x5}
\end{eqnarray}
where the sum is meant to run over all permutations $\Pi$ of 
$\{1,...,n\}$ and thus consist of $n!$ summands.
Recalling the definitions above Eq.~(\ref{a52}), the
last factors $\sigma^+_{l_1} \cdots \sigma^+_{l_n} | 0 \rangle$
in Eq.~(\ref{x4}) may be viewed as ``$n$-spin excitations'',
i.e., a product state where all spins with indices $l_1,...,l_n$
point ``up'' and all other spins point ``down''.
Thus, the energy eigenstate on the left-hand side 
of Eq.~(\ref{x4}) amounts to a weighted sum over all 
possible $n$-spin excitations.
Finally, one readily verifies that 
$\sigma^+_{l_1} \cdots \sigma^+_{l_n} | 0 \rangle$
is an eigenstate of the conserved quantity $S^z$
(total spin) defined below Eq.~(\ref{a1}),
and that the corresponding eigenvalue is given 
by $(2n-L)/2$.
It follows that also Eq.~(\ref{x4}) is an eigenstate of
$S^z$ with eigenvalue $(2n-L)/2$.
In conclusion, every eigenspace of $S^z$ is spanned
by all the $\binom{L}{n}$ energy eigenstates (\ref{x4}) 
with a fixed value of $n$, and is invariant under the 
system dynamics generated by $H$.}

\subsection{Remaining steps}
\label{appA3}

The remaining task is to 
explicitly determine the quantities $U_{kl}$ and $\tilde E_k$ 
appearing in {Eqs.}~(\ref{a49}), (\ref{a50}).
Moreover, we again focus on Hamiltonians (\ref{a48})
of the specific form (\ref{a15}), 
i.e., with
\begin{eqnarray}
B_{kl}= - \delta_{k l+1}J_l/2 - \delta_{k+1 l} J_k/2 - \delta_{kl} \hh_l
\ ,
\label{a66}
\end{eqnarray}
for all $k,l\in\{1,...,L\}$.

Once this problem is solved, one can conclude from
{Eqs.}~(\ref{a6}), (\ref{a56}), and (\ref{a57}) that
\begin{eqnarray}
H \,|\vec b\rangle & = & E(\vec b)\, |\vec b\rangle
\ ,
\label{a67}
\\
E(\vec b) & := & \hh_{tot}+\sum_{l=1}^L b_l\, \tilde E_l
\ .
\label{a68}
\end{eqnarray}
Accordingly, it is natural to employ the eigenbasis $|\vec b\rangle$ of
$H$ for all the remaining calculations. 
To this end, it is in view of {Eq.}~(\ref{a53}) also necessary to express the considered 
observables in terms of the operators $ f^\dagger_l$ and $ f_l$.
For the specific observables in {Eqs.}~(\ref{a12})-(\ref{a14}) this is readily 
achieved by inverting {Eq.}~(\ref{a49}), yielding
\begin{eqnarray}
c_l = \sum_{k=1}^L 
U^\ast_{kl}
\,  f_k
\ ,
\label{a69}
\end{eqnarray}
and then introducing this relation into 
{Eqs.}~(\ref{a9}) and (\ref{a12})-(\ref{a14}).

According to the discussion below  Eq.~(\ref{a48}), we have to solve
the eigenvalue problem $BU^\dagger = U^\dagger D$.
On the level of the matrix elements, this amounts to $L^2$ 
equations of the form
\begin{eqnarray}
\sum_{j} B_{lj}U^\ast_{kj}= \sum_{j} U^\ast_{jl} D_{jk}=U^\ast_{kl} \tilde E_k
\label{a70}
\end{eqnarray}
for arbitrary $k,l\in\{1,...,L\}$.

To simplify the notation,
let us temporarily consider $k\in\{1,...,L\}$ as arbitrary but fixed,
and employ the abbreviations
\begin{eqnarray}
E & := & \tilde E_k
\ ,
\label{a71}
\\
a_l & := & \kappa \, U^\ast_{kl}
\ .
\label{a72}
\end{eqnarray}
Moreover, while the $U_{kl}$ are normalized according to
$\sum_{l}|U_{kl}|^2=1$, we do not require such a normalization 
condition for the $a_l$'s.
This is accounted for by the still arbitrary proportionality
constant $\kappa\in \CC$ in {Eq.}~(\ref{a72}).
Our only requirement is that $\kappa\not=0$, or equivalently, 
that at least one of the $a_l$'s must be non-zero.
Introducing {Eqs.}~(\ref{a66}), (\ref{a71}), and (\ref{a72}) into {Eq.}~(\ref{a70})
then yields
\begin{eqnarray}
 \overline{\delta}_{lL}  \, J_l \,a_{l+1} + \overline{\delta}_{l1}  \, J_{l-1} \, a_{l-1} 
= -2(E+\hh_l) a_l
\ \ 
\label{a73}
\end{eqnarray}
for all $l=1,...,L$, where $\overline{\delta}_{lj}:=1-\delta_{lj}$
is the complementary Kronecker delta (hence 
the quantities
$J_0$, $J_L$, $a_0$, and $a_{L+1}$, which appear in {Eq.}~(\ref{a73}) 
when $l=1$ or $l=L$ but have not been specified until now, 
actually do not matter).

In summary, we are left with
the task to determine $L$ linearly independent
solutions of the $L$ coupled linear equations (\ref{a73}).
The energies $\tilde E_k$ and matrix elements $U_{kl}$ then
readily follow as detailed in and around Eqs.~(\ref{a71}), (\ref{a72}).

{Note that for any given solution $a_l$ of {Eq.}~(\ref{a73}), 
also the real- and imaginary-parts of $a_l$ will be solutions.
It follows that, if it is convenient, we always may assume 
without loss of generality that the $a_l$ and hence 
the $U_{kl}$ are purely real quantities.}

\subsection{Explicit solution for the model from {Eq.}~(\ref{6})}
\label{appA4}

For general models of the form (\ref{a1}), an explicit analytical solution of
the remaining problem from the second-to-last paragraph of the previous subsection 
is still a quite demanding task.
However, for the actual model of interest in our present 
paper, see Eq.~(\ref{6}), the solution is straightforward.
The details are as follows.

Our present special case from  Eq.~(\ref{6}) corresponds to
$J_l=J$ and $\hh_l=\hh $ for all $l$ in {Eq.}~(\ref{a1}),
hence Eq.~(\ref{a73})  takes the form
\begin{eqnarray}
J\,(a_{l+1}+a_{l-1}) & = & - 2 (E+ \hh ) \, a_l \ \, \mbox{for $l=2,...,L-1\,$,}
\ \ \ \ \ \ \ \ 
\label{a74}
\\
J a_2 & = & -2 (E+ \hh) \, a_1 
\ ,
\label{a75}
\\
J a_{L-1} & = & - 2 (E+ \hh ) \, a_L 
\ .
\label{a76}
\end{eqnarray}
One readily verifies by inspection that
the $L$ linearly independent solutions of these equations
(see also  in and around Eqs.~(\ref{a71}), (\ref{a72}))
are given by
\begin{eqnarray}
\tilde E_k & = & - J \cos(k\pi/(L+1)) - \hh
\ ,
\label{a77}
\\
U_{kl} & = & \sqrt{\mbox{$\frac{2}{L+1}$}}\, \sin(lk\pi/(L+1)) =: \tilde S_k^l
\ .
\label{a78}
\end{eqnarray}
In particular, one readily verifies that the normalization 
condition 
$\sum_{k=1}^L (\tilde S_k^l)^2=1$ 
is fulfilled for any 
 $l\in\{1,...,L\}$.
By means of a more involved calculation, 
one also can verify explicitly that, as it must be, 
the orthonormality 
conditions
$\sum_{k=1}^L \tilde S_k^l  \tilde S_k^j = \delta_{jl}$ 
and
$\sum_{k=1}^L \tilde S_l^k  \tilde S_j^k = \delta_{jl}$ 
are fulfilled for all $j,l\in\{1,...,L\}$.

\subsection{Exponential degeneracies of the many-body energies and energy gaps}
\label{appA5}

As pointed out below  Eq.~(\ref{a57}), the set of all many-body energies 
$E_n$ appearing in {Eq.}~(\ref{3}) can be identified with the set of all 
$\tilde E(\vec b)$ in {Eq.}~(\ref{a57}), where $\vec b$ is defined above  Eq.~(\ref{a53}).
[Both sets represent the spectrum of the considered 
many-body Hamiltonian $H$.]
More precisely speaking, 
they are given by the set of all
$E(\vec b)$ in {Eq.}~(\ref{a68}), which differ from the $\tilde E(\vec b)$ by a 
trivial constant, see also {Eqs.}~(\ref{a6}), (\ref{a56}), (\ref{a67}).

To begin with, we show that the Hamiltonian from 
{Eq.}~(\ref{6}) must exhibit many degeneracies
for large $L$, meaning that
$E_n=E_m$ for many $n\not=m$,
or equivalently, $E(\vec b)=E(\vec b')$ 
for many $\vec b\not=\vec b'$.
Let us therefore consider
an arbitrary but fixed $k\in\{1,...,L\}$ 
and define 
$\bar k:=L+1-k$.
Focusing on cases with $k\not = \bar k$ and
observing that $\cos(\pi \bar k/(L+1))=-\cos(\pi k/(L+1))$,
we can conclude from {Eq.}~(\ref{a77}) that
\begin{eqnarray}
\tilde E_k+\tilde E_{\bar k}= - 2\hh
\ .
\label{a88}
\end{eqnarray}
Next we choose two indices $j,k\in\{1,...,L\}$ with
the property that the four numbers $j,\bar j,k,\bar k$ 
are pairwise distinct.
Then, the components of $\vec b=(b_1,...,b_L)$ are chosen 
such that 
$b_j=b_{\bar j}=1$ and 
$b_k=b_{\bar k}=0$,
while all other $s_l$ are arbitrary (but fixed).
Similarly, the components of $\vec b'=(b'_1,...,b'_L)$ are chosen 
such that
$b'_j=b'_{\bar j}=0$,  $b'_k=b'_{\bar k}=1$, and $b'_l=b_l$ for all
other indices $l$.
As a consequence, one readily can infer from {Eqs.}~(\ref{a68})
and (\ref{a88}) that $E(\vec b)=E(\vec b')$.
In other words, each $\vec b$ of the above specified 
structure gives rise to a pair of degenerate energies.
For large $L$ there are very many $\vec b$ of this type, 
hence there must be a large number of degeneracies.
In fact, one readily sees that their number
grows exponentially with the system size $L$.
{We remark that in the special case $G=0$
there also exist considerably simpler examples of 
vectors $\vec b$ and $\vec b'$ with the properties
$\vec b \not = \vec b'$ and $E(\vec b)=E(\vec b')$.}

To arrive at this conclusion, we exploited the result
(\ref{a77}), which in turn is only valid for the specific
Hamiltonians from {Eq.}~(\ref{6}). 
Next, we admit the more general Hamiltonians 
from {Eq.}~(\ref{a1}) and we show that they must 
exhibit many degenerate energy gaps for large $L$,
meaning that $E_m-E_n=E_{m'}-E_{n'}$
for many $m,n,m',n'$ with 
$n\not =m$ and $(m',n')\not = (m,n)$.
Equivalently, we can and will show that 
$E(\vec b)-E(\vec b')=E(\vec r)-E(\vec r')$
for many $\vec b,\vec b',\vec r,\vec r'$ with
$\vec b'\not=\vec b$
and $(\vec r,\vec r')\not=(\vec b,\vec b')$.
Let us therefore
consider any $\vec b:=(b_1,...,b_L)$ with $b_1=1$ and $b_2,...,b_L\in\{0,1\}$
arbitrary but fixed. 
Hence, there are $2^{L-1}$ possible such choices of $\vec b$.
For an arbitrary but fixed such $\vec b$, we choose 
$\vec b':=(b'_1,...,b'_L)$ with $b'_1=0$ and $b'_l=b_l$ for 
$l=2,...,L$.
It follows with {Eq.}~(\ref{a68}) that the energy gap 
$E(\vec b)-E(\vec b')$ equals $E_1$.
Thus, there are $2^{L-1}$ different pairs $(\vec b,\vec b')$ 
with the same energy gap $E(\vec b)-E(\vec b')$.
In other words, the maximal degeneracy of the energy 
gaps is lower bounded by $2^{L-1}$, and thus grows
exponentially with the system size $L$.
(Note that a faster than exponential growth is
impossible since the number of all gaps only 
grows exponentially.)

\section{Dynamical correlation functions}
\label{appB}

Similar results as we will derive them here can 
be found for instance in Refs.~\cite{cru81,sto92,sto95,der00}.
However, the derivation itself is much harder 
to find in the literature.

As in the previous Appendix, 
we consider the generalized XX-model from {Eq.}~(\ref{a1}). 
As in {Eq.}~(\ref{a53}), we define $\vec b:=(b_1,...,b_L)$ 
with 
$b_l\in\{0,1\}$ and  
\begin{eqnarray}
|\vec b\rangle := 
(f_1^\dagger)^{b_1}
(f_2^\dagger)^{b_2}
\cdots 
(f_L^\dagger)^{b_L}
\,  |0\rangle
\ ,
\label{b1}
\end{eqnarray}
implying (see {Eqs.}~(\ref{a67}) and (\ref{a68}))
\begin{eqnarray}
H \,|\vec b\rangle & = & E(\vec b)\, |\vec b\rangle
\ ,
\label{b2}
\\
E(\vec b) & := & \hh_{tot}+\sum_{l=1}^L b_l\, \tilde E_l
\ .
\label{b3}
\end{eqnarray}
Accordingly, it is natural to employ the eigenbasis $|\vec b\rangle$ of
$H$ for all the subsequent calculations. 
To deal with the specific observables in {Eqs.}~(\ref{a9}) and (\ref{a12})-(\ref{a14})
we additionally need the relation (see (\ref{a69}))
\begin{eqnarray}
c_l = \sum_{k=1}^L 
U^\ast_{kl}
\, f_k
\ .
\label{b4}
\end{eqnarray}

As in the main paper, the canonical ensemble is defined as
\begin{eqnarray}
\rho_{can}
& := & Z^{-1} e^{-\beta H}
\ ,
\label{b5}
\\
Z 
& := &
\tr\{e^{-\beta H}\}
\ ,
\label{b6}
\end{eqnarray}
and the corresponding thermal 
expectation 
values are abbreviated as
\begin{eqnarray}
\langle A \rangle_{\rm th}:=\tr\{\rho_{can} A\}
\ .
\label{b7}
\end{eqnarray}

Utilizing the basis $|\vec b\rangle$ to evaluate the traces 
in {Eqs.}~(\ref{b6}) and (\ref{b7}) we thus obtain
\begin{eqnarray}
\langle A \rangle_{\rm th}
& = &
Z^{-1}\sum_{\vec b} e^{-\beta E({\vec b})} \langle {\vec b}|A|{\vec b}\rangle
\ ,
\label{b8}
\\
Z
& = &
\sum_{\vec b} e^{-\beta E({\vec b})}
\ ,
\label{b9}
\end{eqnarray}
where 
\begin{eqnarray}
\sum_{\vec b} := \sum_{b_1=0}^1\sum_{b_2=0}^1\cdots\sum_{b_L=0}^1
\label{b10}
\end{eqnarray}
denotes the sum over all the $2^L$ possible
vectors ${\vec b}:=(b_1,...,b_L)$ with $b_l\in\{0,1\}$.

Finally (and as in the main paper),
\begin{eqnarray}
A(t)
& := &
e^{{iH t}} A e^{{-iHt}}
\label{b11}
\end{eqnarray}
represents the observable $A$ at time $t$ in the Heisenberg picture.

In the following, the same definitions as in {Eqs.}~(\ref{b7}) and (\ref{b11})
will also be adopted for arbitrary (not necessarily Hermitian)
operators $A$.

A particularly important example will turn out to be the (non-Hermitian) 
operator $A=f_l$, yielding with {Eq.}~(\ref{b11}) the relation
\begin{eqnarray}
\dot f_l(t) =
{i\, e^{iH t} [H,f_l] e^{-iHt}}
\ .
\label{b12}
\end{eqnarray}
With {Eqs.}~(\ref{a6}) and (\ref{a50}) we can conclude that
\begin{eqnarray}
[H,f_l] = \sum_{k=1}^L \tilde E_k\, [f^\dagger_k f_k, f_l]
\ .
\label{b13}
\end{eqnarray}
From the basic anti-commutation relations 
(\ref{a51}) one can infer that
$[f^\dagger_k f_k, f_k]=-f_k$ and
$[f^\dagger_k f_k, f_l]=0$ if $k\not=l$,
implying
$[H,f_l]=-\tilde E_l f_l$.
Together with {Eq.}~(\ref{b12}) this yields
$\dot f_l(t) =- i \tilde E_l f_l(t)$
and thus 
\begin{eqnarray}
f_l(t)=
e^{{-i\tilde E_lt}} f_l
\ .
\label{b14}
\end{eqnarray}

\subsection{Evaluation of $\langle \sigma^z_l \sigma^z_j (t)\rangle_{\rm th}$ and $C_{s_l^z s_j^z}(t)$}
\label{appB1}

As a first example, let us consider temporal 
correlations of the form
$\langle \sigma^z_l\sigma^z_j(t)\rangle_{\rm th}$,
where 
$\sigma_l^z =  2c_l^\dagger c_l-1$
according to {Eq.}~(\ref{a14}).
With {Eq.}~(\ref{b11}) it follows that $\sigma_l^z(t)=2c_l(t)^\dagger c_l(t)-1$
and with {Eqs.}~(\ref{b5})-(\ref{b11}) that
$\langle c_l(t)^\dagger c_l(t)\rangle_{\rm th}=\langle c_l^\dagger c_l\rangle_{\rm th}$
independent of $t$.
We thus can conclude that
\begin{eqnarray}
\langle \sigma^z_l \sigma^z_j(t) \rangle_{\rm th}
& = &
4 B(t) + R
\ ,
\label{b15}
\\
B(t) 
& := &
\langle c_l^\dagger c_l  c_j^\dagger(t) c_j(t) \rangle_{\rm th}
\ ,
\label{b16}
\\
R
& := &
1-2\langle c^\dagger_l c_l\rangle_{\rm th}-2 \langle c_j^\dagger c_j\rangle_{\rm th}
\ .
\label{b17}
\end{eqnarray}
Employing {Eq.}~(\ref{b4}) it follows that
\begin{eqnarray}
\langle c^\dagger_l c_l\rangle_{\rm th}
& = &
\sum_{k_1,k_2=1}^L U_{k_1l}U_{k_2l}^\ast \, \langle f_{k_1}^\dagger f_{k_2}  \rangle_{\rm th}
\ .
\label{b18}
\end{eqnarray}
Similarly, {Eqs.}~(\ref{b4}) and (\ref{b14}) imply
\begin{eqnarray}
B(t)
& = &
\sum_{\bf k} U_{k_1l}U_{k_2l}^\ast U_{k_3j}U_{k_4j}^\ast 
e^{{i(\tilde E_{k_3}-\tilde E_{k_4})t}} 
B_{\bf k}
\ ,
\ \ \ \ \ 
\label{b19}
\\
B_{\bf k}
& := &
\langle f_{k_1}^\dagger f_{k_2}  f_{k_3}^\dagger f_{k_4}\rangle_{\rm th}
\ ,
\label{b20}
\end{eqnarray}
where ${\bf k}:=(k_1,k_2,k_3,k_4)$ and the sum in {Eq.}~(\ref{b19}) runs over all
$k_{1,...,4}\in\{1,...,L\}$.

The last term in {Eq.}~(\ref{b18}) can be rewritten by means of {Eq.}~(\ref{b8}) as
\begin{eqnarray}
\langle f_{k_1}^\dagger f_{k_2}  \rangle_{\rm th}
=
Z^{-1}\sum_{\vec b} e^{-\beta E(\vec b)} \langle \vec b| f_{k_1}^\dagger f_{k_2} |\vec b\rangle
\ .
\label{b21}
\end{eqnarray}
Taking into account the basic relations (\ref{a51}), (\ref{a52}), and (\ref{b1}),
a straightforward but somewhat tedious calculation shows that 
$\langle \vec b| f_{k_1}^\dagger f_{k_2} |\vec b\rangle$ must vanish unless 
$k_1=k_2$.

The following alternative line of reasoning is intuitively more appealing
(but actually working out all the details is about equally laborious):
From the discussion below  Eq.~(\ref{a72}) 
it follows that we still have the freedom to replace
all $U_{kl}$ by $e^{i\alpha_k} U_{kl}$, where 
$\alpha_k$ are arbitrary (real-valued) phases,
without changing any physical properties of the system, 
for instance the left-hand side of Eq.~(\ref{b18}).
Moreover, it can be shown that also the quantities 
$\langle f_{k_1}^\dagger f_{k_2}  \rangle_{\rm th}$ on the right-hand side
of {Eq.}~(\ref{b18}) must be independent of those phases.
The latter claim is obvious as far as the quantities $Z^{-1}$ and $e^{-\beta E(\vec b)}$
in {Eq.}~(\ref{b8}) are concerned.
With respect to the remaining
quantities $\langle \vec b | f_{k_1}^\dagger f_{k_2} | \vec b \rangle$ our claim follows 
from the fact that neither in the definition of $|\vec b\rangle$ from {Eq.}~(\ref{b1}) 
nor in the basic anti-commutation relations in {Eq.}~(\ref{a51}) 
those phases play any role.
It is thus quite reasonable to expect that the independence from
all those extra factors $e^{i(\alpha_{k_1}-\alpha_{k_2})}$ 
on the right-hand side {Eq.}~(\ref{b18})
implies that all summands with $k_1\not =k_2$ must be zero.
A more formal verification follows by exploiting that all mixed 
second derivatives $\partial^2/\partial\alpha_{k_1}\partial\alpha_{k_2}$ 
of Eq.~(\ref{b18}) must vanish.

From {Eqs.}~(\ref{a55}) and (\ref{a54}) it moreover follows that
$\langle {\vec b}| f_{k}^\dagger f_{k} |{\vec b}\rangle=\langle {\vec b}| n_k |{\vec b}\rangle=b_k$.
Together with {Eq.}~(\ref{b9}) we thus can rewrite {Eq.}~(\ref{b18}) as
\begin{eqnarray}
\langle c^\dagger_l c_l\rangle_{\rm th}
& = &
\sum_{k=1}^L |U_{kl}|^2 p_k
\ ,
\label{b22}
\\
p_k
& := &
\frac{\sum_{\vec b} b_k\, e^{-\beta E({\vec b})}}{\sum_{\vec b} e^{-\beta E({\vec b})}}
\ .
\label{b23}
\end{eqnarray}
Exploiting {Eqs.}~(\ref{b3}) and (\ref{b10}), the denominator in {Eq.}~(\ref{b23}) can be rewritten as
\begin{eqnarray}
\sum_{\vec b} e^{-\beta E({\vec b})}
& = & 
e^{-\beta \hh_{tot}}\prod_{l=1}^L z_l
\ ,
\label{b24}
\\
z_l 
& := &
\sum_{b_l=0}^1 e^{-\beta b_l \tilde E_l}
\ .
\label{b25}
\end{eqnarray}
Likewise, the numerator in {Eq.}~(\ref{b23}) can be written in almost the same form
as in {Eq.}~(\ref{b24}), except that the factor $z_k$ must now be replaced by
\begin{eqnarray}
\tilde z_k 
& := &
\sum_{b_k=0}^1 b_k \, e^{-\beta b_k \tilde E_k}
\ .
\label{b26}
\end{eqnarray}
Altogether, this yields
\begin{eqnarray}
p_k 
=
\frac{\tilde z_k}{z_k}
= 
\frac{\sum_{b_k=0}^1 b_k\, e^{-\beta b_k \tilde E_k}}{\sum_{b_k=0}^1 e^{-\beta b_k \tilde E_k}}
= 
\frac{e^{-\beta \tilde E_k}}{1+e^{-\beta \tilde E_k}}
\ .
\label{b27}
\end{eqnarray}
Hence, {Eq.}~(\ref{b22}) can be rewritten as
\begin{eqnarray}
\langle c^\dagger_l c_l\rangle_{\rm th}
=
\sum_{k=1}^L |U_{kl}|^2 f(\tilde E_k)
\ ,
\label{b28}
\end{eqnarray}
where
\begin{eqnarray}
f(E):=\frac{1}{e^{\beta E}+1}
\label{b29}
\end{eqnarray}
is the Fermi function.

The evaluation of {Eqs.}~(\ref{b19}) and (\ref{b20}) is in principle similar but in detail more
involved. In particular,  {Eq.}~(\ref{b20}) is again found to vanish
unless each of the two creation operators has a ``partner'' 
with identical index among the two annihilation operators
(see also the intuitive arguments below  Eq.~(\ref{b21})).
More precisely speaking, there remain three types of 
possibly non-vanishing contributions.
(i) $k_1=k_2=k_3=k_4$. 
(ii) $k_1=k_2$ and $k_3=k_4$ and $k_1\not = k_3$
(or, equivalently, $k_1\not = k_4$).
(iii) $k_1=k_4$ and $k_2=k_3$ and $k_1\not = k_2$.
{Note that} the extra conditions $k_1\not = k_3$ in (ii) and
$k_1\not = k_2$ in (iii) are needed to make sure that
each possibly non-vanishing contribution is contained 
in one and only one of the three subsets (i)-(iii).

The further steps are similar as in {Eqs.}~(\ref{b22})-(\ref{b29})
and therefore not explained in detail.
In case (i) one finds (setting $k:=k_1$) that 
\begin{eqnarray}
\langle {\vec b}|f_{k}^\dagger f_{k}  f_{k}^\dagger f_{k}|{\vec b}\rangle
=
\langle {\vec b}|n_k^2|{\vec b}\rangle 
=
b_k^2=b_k
\label{b30}
\end{eqnarray}
and thus 
$B_{\bf k}=f(\tilde E_k)$.
In case (ii) one finds that 
\begin{eqnarray}
\langle {\vec b}|f_{k_1}^\dagger f_{k_1}  f_{k_3}^\dagger f_{k_3}|{\vec b}\rangle
=
\langle {\vec b}|n_{k_1} n_{k_3} |{\vec b}\rangle
=
b_{k_1}b_{k_3}
\label{b31}
\end{eqnarray}
and thus 
$B_{\bf k}=f(\tilde E_{k_1})f(\tilde E_{k_3})$.
In case (iii) one finds that 
\begin{eqnarray}
& & 
\langle {\vec b}|f_{k_1}^\dagger f_{k_2}  f_{k_2}^\dagger f_{k_1}|{\vec b}\rangle
=
\langle {\vec b}|f_{k_1}^\dagger f_{k_1}(1-f_{k_2} ^\dagger f_{k_2}) |{\vec b}\rangle
\nonumber
\\
& & 
=
\langle {\vec b}|n_{k_1}(1-n_{k_2}) |{\vec b}\rangle
=b_{k_1}(1-b_{k_2})
\label{b32}
\end{eqnarray}
and thus 
$B_{\bf k}=f(\tilde E_{k_1})[1-f(\tilde E_{k_2})]$.

Accordingly, {Eq.}~(\ref{b19}) can be rewritten as
\begin{eqnarray}
B(t) & = &  B^{(i)}+B^{(ii)}+B^{(iii)}(t)
\ ,
\label{b33}
\\
B^{(i)}
& := & 
\sum_{k=1}^L |U_{kl}|^2|U_{kj}|^2 f(\tilde E_k)
\ ,
\label{b34}
\\
B^{(ii)}
& := & 
\sum_{k_1,k_3=1}^L |U_{k_1l}|^2|U_{k_3j}|^2 f(\tilde E_{k_1}) f(\tilde E_{k_3})
\nonumber
\\
& & 
- \sum_{k=1}^L |U_{kl}|^2|U_{kj}|^2 [f(\tilde E_{k})]^2 
\ ,
\label{b35}
\\
B^{(iii)}(t)
& := & 
\sum_{k_1,k_2=1}^L 
U_{k_1l}U_{k_2l}^\ast U_{k_2j}U_{k_1j}^\ast
\, e^{{i(\tilde E_{k_2}-\tilde E_{k_1})t}} 
\nonumber
\\
& & 
\ \ \times\ f(\tilde E_{k_1}) 
[1-f(\tilde E_{k_2})]
\nonumber
\\
& & 
- \sum_{k=1}^L |U_{kl}|^2|U_{kj}|^2 f(\tilde E_{k})
[1-f(\tilde E_k)]
\ . \ \ \ \
\label{b36}
\end{eqnarray}
The last sum in {Eq.}~(\ref{b35}) has its origin in the fact that $k_1=k_3$ has been
excluded in case (ii), and likewise for the last sum in {Eq.}~(\ref{b36}).

Our first observation is that the contributions to {Eqs.}~(\ref{b33})
by {Eq.}~(\ref{b34}) and by the last sums in {Eqs.}~(\ref{b35}) and (\ref{b36})
cancel each other, i.e., only the first sums in {Eqs.}~(\ref{b35}) and (\ref{b36})
remain.
Second, the first sum in {Eq.}~(\ref{b35}) can be rewritten by means of {Eq.}~(\ref{b28})
as $\langle c^\dagger_l c_l\rangle_{\rm th}\langle c^\dagger_j c_j\rangle_{\rm th}$.
Finally, in the first sum in {Eq.}~(\ref{b36}) we can exploit that 
\begin{eqnarray}
1-f(E)=f(-E)
\label{b37}
\end{eqnarray}
according to {Eq.}~(\ref{b29}).
Altogether, we thus obtain
\begin{eqnarray}
B(t) 
& = &  
\langle c^\dagger_l c_l\rangle_{\rm th}\langle c^\dagger_j c_j\rangle_{\rm th}
+{\cal B}^+_{lj}(t) [{\cal B}^-_{lj}(t)]^\ast
\ ,
\label{b38}
\\
{{\cal B}^\pm_{lj}(t)}
& := & 
\sum_{k=1}^L U_{kl}U_{kj}^\ast  
e^{{-i\tilde E_k t}} 
f(\pm \tilde E_{k})
\ .
\label{b39}
\end{eqnarray}

Since $\sigma_l^z =  2c_l^\dagger c_l-1$ (see below Eq.~(\ref{b14}))
and $s_l^a=\sigma_l^a/2$ (see paragraph before {Eq.}~(\ref{a2}))
we can infer from {Eq.}~(\ref{b28}) that
\begin{eqnarray}
\langle s_l^z \rangle_{\rm th}
=
\sum_{k=1}^L |U_{kl}|^2 f(\tilde E_k) - 1/2
\label{b40}
\end{eqnarray}
for all $l\in\{1,...,L\}$.
By exploiting the definition of $f(E)$ in {Eq.}~(\ref{b29}) and the
property $\sum_{l=1}^L |U_{jl}|^2 = 1$ (see below Eq.~(\ref{a78}))
this yields
\begin{eqnarray}
\langle s_l^z \rangle_{\rm th}
= \frac{1}{2}\sum_{k=1}^L |U_{kl}|^2 \, \tanh(-\beta \tilde E_k/2)
\ .
\label{b41}
\end{eqnarray}

Likewise, upon adopting the definition (\ref{12}) for the dynamical correlation 
function $C_{\! V\!\!A}(t)$ of two arbitrary operators $V$ and $A$ 
we can conclude from {Eqs.}~(\ref{b15})-(\ref{b17}) and (\ref{b38}) that
\begin{eqnarray}
C_{s_l^z s_j^z}(t) = {\cal B}^+_{lj}(t) [{\cal B}^-_{lj}(t)]^\ast
\label{b42}
\end{eqnarray}
for all $l,j\in\{1,...,L\}$.

\subsection{Evaluation of more general temporal correlations}
\label{appB2}

As a second example, we consider the observables $\sigma_l^x$, see
Eqs.~(\ref{a12}), (\ref{a16}), and (\ref{b4}).
Similarly as above  Eq.~(\ref{b22}) and around  Eq.~(\ref{b30}),
one finds after a few steps that we
have to evaluate sums of expressions of the form 
$\langle \vec b|O|\vec b\rangle$,  where $O$ is always a product
of an odd number of creation and annihilation operators.
As a consequence 
(see also the intuitive arguments below  Eq.~(\ref{b21})), 
all those expressions $\langle \vec b|O|\vec b\rangle$ 
must be zero,
implying $\langle \sigma_l^x \rangle_{\rm th}=0$.
Analogous conclusions apply to $\sigma_l^y$, and hence
\begin{eqnarray}
\langle s_l^{x,y} \rangle_{\rm th}
=
0
\label{b43}
\end{eqnarray}
for all $l\in\{1,...,L\}$.
The same result can also be derived by means of symmetry considerations,
see the Supplemental Material of Ref.~\cite{eid23}.

Analogously one finds that temporal 
correlations of the form
$\langle\sigma_l^z \sigma_j^x(t) \rangle_{\rm th}$
give rise exclusively to odd numbers of creation and 
annihilation operators, resulting in
\begin{eqnarray}
C_{s_l^z s_j^{x,y}}(t) = 0
\label{b44}
\end{eqnarray}
for all $l,j\in\{1,...,L\}$.

Finally, temporal correlations of the form
$\langle\sigma_l^a \sigma_j^b(t) \rangle_{\rm th}$
with $a,b\in\{x,y\}$ turn out to be in general 
considerably more difficult to evaluate.
The reason is the factor $Z_l$ from {Eqs.}~(\ref{a9}) or (\ref{a16})
appearing in {Eqs.}~(\ref{a12}) and (\ref{a13}).
An exception is the case $l=j=1$ 
(since $Z_1=1$).
Omitting the details, the final result is
\begin{eqnarray}
C_{s_1^x s_1^x}(t) 
& = & 
C_{s_1^y s_1^y}(t)
=
\frac{{\cal B}^+_{11}(t) + [{\cal B}^-_{11}(t)]^\ast}{4}
\ ,
\label{b45}
\\
C_{s_1^y s_1^x}(t) 
& = & 
- C_{s_1^x s_1^y}(t)
=
\frac{{\cal B}^+_{11}(t) - [{\cal B}^-_{11}(t)]^\ast}{4i}
\ ,
\label{b46}
\end{eqnarray}
where ${\cal B}^\pm_{11}(t)$ is defined in {Eq.}~(\ref{b39}),
i.e.,
\begin{eqnarray}
{\cal B}^\pm_{11}(t)
& = & 
\sum_{k=1}^L |U_{k1}|^2  
e^{{-i\tilde E_k t}} 
f(\pm \tilde E_{k})
\ .
\label{b47}
\end{eqnarray}

\section{Explicit evaluation of Eq.~(\ref{13}) for specific examples}
\label{appC}

Our first examples are perturbations $V$ and observables $A$ 
of the form
\begin{eqnarray}
V=s_\nu^z \ ,\ \ 
A=s_\alpha^z
\label{c1}
\end{eqnarray}
with $\nu,\alpha\in\{1,..,L\}$ arbitrary but fixed.
Exploiting 
{Eqs.}~(\ref{b42}) and (\ref{b39})
it follows that
\begin{eqnarray}
C_{s_\nu^z s_\alpha^z}(t) 
& = &
\sum_{m,n=1}^L U_{mn}^{\nu\alpha} \, \tilde f_{mn} \, 
e_{mn}(t)
\ ,
\label{c2}
\\
U_{mn}^{\nu\alpha}
& := &
U_{m\nu} U_{m\alpha}^\ast  U_{n\nu}^\ast  U_{n\alpha}  
\ ,
\ \ \
\label{c3}
\\
\tilde f_{mn}
& := &
f(\tilde E_{m})  f(-\tilde E_{n})
\ ,
\ \ \
\label{c4}
\\
e_{mn}(t)
& := &
e^{{-i(\tilde E_{m}-\tilde E_{n})t}}
\ ,
\label{c5}
\end{eqnarray}
and with {Eq.}~(\ref{13}) that
\begin{eqnarray}
\langle s_\alpha^z \rangle_t - \langle s_\alpha^z\rangle _{\rm th}
& = & 
g \beta \!
\sum_{m,n=1}^L U_{mn}^{\nu\alpha}\, f_{mn}
\, e_{mn}(t)
\ ,
\ \ \ \ \ \ 
\label{c6}
\\
f_{mn}
& := &
\tilde f_{mn}\, Q_{mn}
\ ,
\label{c7}
\\
Q_{mn}
& := &
\sum_{k=0}^\infty  \frac{(\beta(\tilde E_m-\tilde E_n))^k}{(k+1)!}
\ .
\label{c8}
\end{eqnarray}
We thus can conclude that
\begin{eqnarray}
Q_{mn}
& = & 
F(\beta(\tilde E_m-\tilde E_n))
\ ,
\label{c9}
\\
F(x)
& := & \sum_{k=0}^\infty  \frac{x^k}{(k+1)!}= \frac{e^x-1}{x}
\ .
\label{c10}
\end{eqnarray}
Furthermore, the quantity $f_{mn}$ from {Eq.}~(\ref{c7})
can be rewritten by means of 
{Eqs.}~(\ref{b29}), (\ref{c4}), (\ref{c9}), (\ref{c10})
as
\begin{eqnarray}
f_{mn}
& = & 
\frac{\sinh\left(\beta\frac{\tilde E_m-\tilde E_n}{2}\right)}{2\beta(\tilde E_m-\tilde E_n)}
\,
\frac{1}{\cosh(\beta \frac{\tilde E_m}{2})\cosh(\beta  \frac{\tilde E_n}{2})}
\nonumber
\\
& = &
\frac{\tanh(\beta \frac{\tilde E_m}{2})-\tanh(\beta \frac{\tilde E_n}{2})}{2\beta(\tilde E_m-\tilde E_n)}
\ .
\label{c11}
\end{eqnarray}
A more detailed discussion 
of this quantity is provided
in Appendix \ref{appD}.
Introducing {Eqs.}~(\ref{a77}) and (\ref{a78}) into these {results}, 
one readily recovers {Eq.}~(\ref{15}).

{In passing we note that the following properties of {Eq.}~(\ref{c6}),
which we also mentioned in the main text for the specific model 
considered therein,
are in fact generally valid.
The starting point is the observation at the end of Sec.~\ref{appA4}
that all $U_{kl}$ may be assumed to be real-valued.
It follows that the $U_{mn}^{\nu\alpha}$ in {Eq.}~(\ref{c3}) are real-valued, implying that
{Eq.}~(\ref{c6}) is symmetric with respect to the indices $\nu$ and $\alpha$.
Moreover, {Eq.}~(\ref{c3}) as well as {Eq.}~(\ref{c11}) are symmetric with respect
to $m$ and $n$. Upon interchanging the summation indices $m$ and $n$
in {Eq.}~(\ref{c6}) one thus can conclude that the entire sum must
be real-valued, and that it must be time-inversion invariant.
Finally, {Eq.}~(\ref{c11}) is obviously an even function of $\beta$,
hence the same property is inherited by {Eq.}~(\ref{c6}).}

As our next examples we consider perturbations 
$V$ and observables $A$ of the form $V=A=s_1^a$ 
with $a\in\{x,y\}$.
In this case, upon exploiting {Eqs.}~(\ref{b45}) and (\ref{b47})  
one readily arrives by similar (but simpler) 
calculations as above at the result (\ref{76}) in the main paper.

Likewise, for examples of the form $V=s_1^y$ and $A=s_1^x$ 
one may exploit {Eq.}~(\ref{b46}) to recover the result 
(\ref{77}).
Finally, the case $V=s_1^x$ and $A=s_1^y$
simply amounts to a sign change compared to
the former case according to {Eq.}~(\ref{b46}).

\section{Properties of $ f_{mn}$}
\label{appD}

We rewrite {Eq.}~(\ref{17}) as
\begin{eqnarray}
 f_{mn} 
& = &
g(\beta(\tilde E_m-\tilde E_n)/2,\beta \tilde E_n/2)/4
\ ,
\label{d1}
\\
g(x,y)
& := &
\frac{\tanh(x+y)-\tanh(y)}{x}
\ .
\label{d2}
\end{eqnarray}
Considering $y$ as arbitrary but fixed, Taylor's theorem implies
\begin{eqnarray}
\tanh(x+y) = \tanh(y) + x\, \tanh'(\xi)
\label{d3}
\end{eqnarray}
for some $\xi$ which in general depends on $x$ and $y$ but which is
known to always take a value between $y$ and $x+y$.
Since $\tanh'(\xi)=1/\cosh^2(\xi)\in (0,1]$ for all $\xi\in\RR$ it follows by
introducing {Eq.}~(\ref{d3}) into {Eq.}~(\ref{d2}) that
$0 < g(x,y)\leq 1$
for all $x,y\in\RR$ and with {Eq.}~(\ref{d1}) that
\begin{eqnarray}
0 <  f_{mn}\leq 1/4
\label{d4}
\end{eqnarray}
for all $m,n$.
Furthermore, if $|x|\ll 1$ it follows that
$g(x,y)\simeq 1/\cosh^2(y)$, implying the approximation
\begin{eqnarray}
 f_{mn}\simeq \frac{1}{4\, \cosh^2(\beta \tilde E_n/2)}
\ \ \mbox{if $|\beta(\tilde E_m-\tilde E_n)|\ll1$.}
\ \ 
\label{d5}
\end{eqnarray}
Likewise, considering the limit $x\to 0$ one can infer (without any approximation) 
that
\begin{eqnarray}
 f_{nn} = \frac{1}{4\, \cosh^2(\beta \tilde E_n/2)}
\label{d6}
\end{eqnarray}
for all $n$.

While all so far considerations are valid for general models of the form (\ref{a1}),
let us finally focus on the model of actual interest in the main paper,
see {Eq.}~(\ref{6}), for which the energies $\tilde E_n$ are given by 
Eq.~(\ref{19}). 
As in Sec.~\ref{s43} we furthermore focus on the case
\begin{eqnarray}
|\beta J|\ll 1
\ .
\label{d7}
\end{eqnarray}
In view of Eq.~(\ref{19}) 
it follows that $|\beta(\tilde E_m-\tilde E_n)| \ll 1$
for all $m,n$, and that $\cosh(\beta \tilde E_n/2)\simeq \cosh(\beta \hh h/2)$
for all $n$. Together with {Eq.}~(\ref{d5}) we thus arrive at the approximation
\begin{eqnarray}
 f_{mn}\simeq \frac{1}{4\, \cosh^2(\beta \hh /2)}
\label{d8}
\end{eqnarray}
for all $m,n$.
Moreover, it is straightforward to verify that the relative error of this 
approximation is at most of the order of $|\beta J|$.

\section{Evaluation and discussion of $\langle s_l^z \rangle _{\rm th}$}
\label{appE}

While most of the considerations in the previous Appendices \ref{appA}-\ref{appD}
are valid for general models of the form (\ref{a1}),
we will focus in this Appendix on the specific XX-model 
from Eq.~(\ref{6}) of the main paper.
Hence, we can utilize the explicit solutions 
obtained in Sec.~\ref{appA4}.

Recalling the property $\sum_{l=1}^L |U_{jl}|^2 = 1$ 
(see below Eq.~(\ref{a78})) we can infer from {Eq.}~(\ref{b41})
that (as expected)
\begin{eqnarray}
\langle s_l^z \rangle_{\rm th}
\in [-1/2,1/2]
\ .
\label{e1}
\end{eqnarray}
Furthermore, by exploiting {Eqs.}~(\ref{a77}) and (\ref{a78}) we can rewrite 
{Eq.}~(\ref{b41}) as
\begin{eqnarray}
\langle s_l^z \rangle_{\rm th}
& = &
F(\beta J/2,\beta \hh /2)/(L+1)
\ ,
\label{e2}
\\
F(x,y)
& := &
\sum_{k=1}^L 
\sin^2({\varphi}_k l)
\, \tanh( x \cos({\varphi}_k) + y )
\ ,
\label{e3}
\\
{\varphi}_k
& := &
\pi k/(L+1)
\ .
\label{e4}
\end{eqnarray}
Similarly as below Eq.~(\ref{22}),
upon replacing all indices $k$ {in Eq. (\ref{e3})}
by $L+1-k$ one can
infer that $F(x,y)=F(-x,y)$.
Furthermore, it is obvious that $F(-x,-y)=-F(x,y)$,
and, by combining both relations, that
$F(x,-y)=-F(x,y)$.

Altogether we thus can conclude that
$\langle s_l^z \rangle_{\rm th}$ 
in {Eq.}~(\ref{e1}) must be an even 
function of $J$, and an odd function of $\beta $
and of $\hh $.
In particular,
\begin{eqnarray}
\langle s_l^z \rangle_{\rm th}
=
0
\ \mbox {if $\hh =0$ or $\beta=0$.}
\label{e5}
\end{eqnarray}

Turning to asymptotically large $\hh$, and assuming that $\beta >0$,
Eq.~(\ref{a77}) implies $\beta \tilde E_k\to- \infty$.
Together with {Eq.}~(\ref{b41}) it follows that
\begin{eqnarray}
\langle s_l^z \rangle_{\rm th} \to 1/2
\ \mbox {if $\beta>0$ and $\hh \to\infty$,}
\label{e6}
\end{eqnarray}
and symmetrically (see above) for $\hh \to-\infty$ 
or $\beta<0$.

Finally, we focus on the case
\begin{eqnarray}
|\beta J|\ll 1
\ .
\label{e7}
\end{eqnarray}
Expanding {Eq.}~(\ref{e3}) up to the second order in $x$,
a straightforward but somewhat lengthy calculation then 
yields 
\begin{eqnarray}
\langle s_l^z \rangle_{\rm th} 
& = &
\frac{1}{2} \tanh(\mbox{$\frac{\beta \hh }{2}$})
\left[
1 - (\beta J)^2\gamma
+\ord((\beta J)^4)
\right],
\ \ \ \ \ \ \ 
\label{e8}
\\
\gamma
& := &
\frac{L}{L+1}
\frac{2-\delta_{l1}-\delta_{lL}}
{4\,\cosh^2(\mbox{$\frac{\beta \hh }{2}$})}
\ ,
\label{e9}
\end{eqnarray}
where $\delta_{lj}$ is the Kronecker delta.
Most notably, the sites with $l=1$ and $l=L$ thus exhibit in general 
a different behavior than the rest of the chain.
It seems reasonable to expect that also the sites with $l=2$ and $l=L-1$
will exhibit such differences in fourth order of $\beta J$, and so on.

\section{Some properties of  $J_k(t)$}
\label{appF}

\begin{figure*}
\hspace*{-0.4cm}
\includegraphics[scale=1.05]{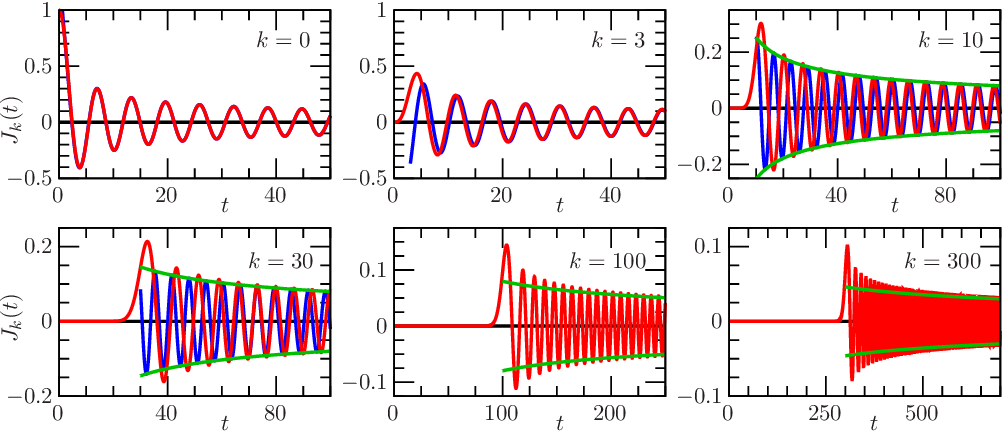}
\caption{
Red: The Bessel functions $J_k(t)$ for
$k=0,3,10,30,100,300$.
Note the different scalings of the axes.
Blue: the large-$t$ asymptotics from {Eq.}~(\ref{f8})
for $t\geq k$. 
Green: The envelopes $\pm\sqrt{2\pi/t}$ 
of the damped (blue) oscillations from {Eq.}~(\ref{f8}). 
For $k=0$ the blue curve ist almost 
completely covered by the red curve.
With increasing $k$, 
the convergence of the blue towards the red curves 
takes longer and longer,
in agreement with the convergence criterion 
below Eq.~(\ref{f8}).
For $k=30$ this convergence is already 
far outside the plotted range.
For $k=100$ and $k=300$ the convergence
is reached even much later, hence the blue 
curves have been omitted altogether.
On the other hand, the envelopes
of the damped red and blue oscillations
converge much earlier than the 
frequencies for large $k$.
}
\label{fig12}
\end{figure*}

Focusing on Bessel functions of the first kind $J_k(t)$ 
with $k\in\NN_0$, their series expansion assumes the form
(see 
(9.1.10) in \cite{abram})
\begin{eqnarray}
J_k(t) 
& = & 
(t/2)^k \sum_{n=0}^\infty \frac{(-t^2/4)^n}{n! (n+k)!}
\ .
\label{f1}
\end{eqnarray}
Approximations for small $t$ can thus be readily inferred.

For sufficiently large $t$ it is known
(see (9.2.5), (9.2.8)-(9.2.10) in \cite{abram})
 that 
\begin{eqnarray}
J_k(t) 
& = &
\sqrt{\frac{2}{\pi t}}
\big[ q_k(t) \cos (t-\phi_k)  -  r_k(t) \sin (t-\phi_k) \big]
,
\ \ \ \ \
\label{f2}
\\
\phi_k
& := & 
\pi k/2+\pi/4
\ ,
\\
\label{f3}
q_k(t)
& := &
\sum_{n=0}^\infty (-1)^n \frac{c_k(2n)}{t^{2n}}
\ ,
\label{f4}
\\
r_k(t)
& := &
\sum_{n=0}^\infty (-1)^n \frac{c_k(2n+1)}{t^{2n+1}}
\ ,
\label{f5}
\\
c_k(0)
& := &
1
\ ,
\label{f6}
\\
c_k(n)
& := &
\frac{(\kappa-1^2)(\kappa-3^2)\cdots (\kappa-(2n-1)^2)}{8^n\, n!}
\ ,
\label{f7}
\end{eqnarray}
where $\kappa:=4k^2$ and $n\geq 1$ in {Eq.}~(\ref{f7}).
More precisely speaking, the remainder after $n$ terms in the expansions
(\ref{f4}), (\ref{f5}) does not exceed the $n+1$-th term in 
absolute value.
It follows that 
\begin{eqnarray}
J_k(t)\simeq \sqrt{\frac{2}{\pi t}}\, \cos (t-\phi_k) 
\label{f8}
\end{eqnarray}
amounts to a good approximation 
at least
for 
$t \gg 
|k^2-1/4|$
(see also below Eq.~(58) in \cite{wolf}),
and 
\begin{eqnarray}
J_k(t)\simeq \sqrt{\frac{2}{\pi t}}\,\big[\!\cos (t-\phi_k) - 
\frac{4k^2-1}{8\, t} \sin (t-\phi_k)]
\ \ \
\label{f9}
\end{eqnarray}
to an even better approximation 
at least for $t \gg |(k^2-1/4)(k^2-9/4)|$,
and so on.

For an arbitrary but fixed $t$ it is known
(see  (9.3.1) in \cite{abram})
that
\begin{eqnarray}
J_k(t)\sim  \frac{1}{\sqrt{2\pi k}}\left( \frac{e t}{2k}\right)^k
\ \mbox{for $k\to\infty$.}
\label{f10}
\end{eqnarray}

Finally, we derive yet another interesting property which seems not
generally known.
Focusing on $t\geq 0$ and
observing that $(n+k)!\geq n!\,k!$ 
we can conclude 
from {Eq.}~(\ref{f1}) that
\begin{eqnarray}
|J_k(t)| 
& \leq & \frac{(t/2)^k}{k!} D_k
\ ,
\label{f11}
\\
D_k & := &
\sum_{n=0}^\infty \left(\frac{(t/2)^{n}}{n!}\right)^2
\ .
\label{f12}
\end{eqnarray}
Since any sum $\sum_{n=1}^N c_n$ with non-negative summands $c_n$
satisfies the inequality $(\sum_{n=1}^N c_n)^2\geq \sum_{n=1}^N c_n^2$
we obtain
\begin{eqnarray}
D_k\leq \left(\sum_{n=0}^\infty \frac{(t/2)^{n}}{n!}\right)^2= \left( e^{t/2}\right)^2 = e^t
\ .
\label{f13}
\end{eqnarray}
Moreover, $k!$ in {Eq.}~(\ref{f11}) can be lower bounded by $(k/e)^k$ according 
to some suitable version of Stirling's formula, yielding
\begin{eqnarray}
|J_k(t)| 
& \leq & \left( \frac{t e}{2k}\right)^k e^{t}
=
e^{t- k\, q}
\ ,
\label{f14}
\\
q 
& := & 
\ln k - \ln t +  \ln2 - 1
\ .
\label{f15}
\end{eqnarray}
Focusing on $t\leq k/2$ it follows that $\ln t\leq\ln k -\ln 2$,
hence
\begin{eqnarray}
q\geq 2\ln 2 - 1 \geq 0.38
\ .
\label{f16}
\end{eqnarray}
For any given $a>0$ we thus can conclude that
\begin{eqnarray}
|J_k(t)|
& \leq &
e^{-a}
\ \mbox{if $t\leq 0.38\, k - a$}
\ .
\label{f17}
\end{eqnarray}
Note that our previous assumption $t\leq k/2$ is 
automatically fulfilled.

The main message of {Eq.}~(\ref{f17}) is that the Bessel 
functions $J_k(t)$ remain
negligibly small up to times $t$ 
which grow linearly with $k$. 
The quantitative numerical examples in
Fig.~\ref{fig12} suggest that this seems indeed 
to be the case up to times $t$ close to $k$, 
i.e., the analytical upper bound in {Eq.}~(\ref{f17}) 
is still not very tight.
Fig.~\ref{fig12} moreover suggests that 
this initial near-zero phase for $t<k$
directly goes over into 
damped oscillations for $t>k$, 
whose envelopes 
are right away in quite good agreement 
with those of the analytical long-time 
asymptotics from {Eq.}~(\ref{f8}),
while the numerically observed 
frequencies of the oscillations
remain notably slower than in {Eq.}~(\ref{f8})
up to considerably larger times,
in agreement with the convergence 
criterion below Eq.~(\ref{f8}).

\section{Derivation of Eq.~(\ref{69})}
\label{appG}

Exploiting {Eqs.}~(\ref{65})-(\ref{67}) it follows that the 
integrand in {Eq.}~(\ref{64}),
\begin{eqnarray}
I(x):=  - 
\cos(k x) \tanh(\beta |J| \cos(x)/2) \exp\{i t \cos(x)\}
, 
\ \ \ \ \ \ 
\label{g1}
\end{eqnarray}
is an even and $2\pi$-periodic function of $x$ 
with the property
$I(x+\pi)=- I^\ast(x)$.
We thus can conclude from {Eq.}~(\ref{64}) that
\begin{eqnarray}
\tilde B_0(t)
& = &
\int\limits_{-\pi}^\pi dx  \frac{I(x)}{2\pi} 
= 
\int\limits_{-\pi/2}^{3\pi/2} \!dx  \frac{I(x)}{2\pi} 
=\frac{K(t)+\tilde K(t)}{2\pi},
\ \ \ \ \ \ 
\label{g2}
\\
K(t)
& := &
\int_{-\pi/2}^{\pi/2} dx  \, I(x)
\ ,
\label{g3}
\\
\tilde K(t)
& := &
\int_{-\pi/2}^{\pi/2} dx \,  I(x+\pi)
= - K^\ast(t)
\ ,
\label{g4}
\end{eqnarray}
implying
\begin{eqnarray}
\Im(\tilde B_0(t)) = \Im(K(t))/\pi
\ .
\label{g5}
\end{eqnarray}

Next, we go over from the integration variable 
$x$ in {Eq.}~(\ref{g3}) to $y:=2\sin(x/2)$
(and we tacitly restrict ourselves to the integration 
domain on the right-hand side, i.e. to $x\in[-\pi/2,\pi/2]$).
It follows that $dy/dx=\cos(x/2)=\sqrt{1-\sin^2(x/2)}$,
where we exploited that $\cos(x/2)\geq 0$ for all 
$x\in[-\pi/2,/\pi/2]$.
Hence, $dx = dy (1-y^2/4)^{-1/2}$.
Moreover $\cos(x)=1-2\sin^2(x/2)=1-y^2/2$.
Altogether, we thus obtain
\begin{eqnarray}
K (t)
& = &
\int_{-\sqrt{2}}^{\sqrt{2}} dy\,
g(y)
\exp\{it(1-y^2/2)\}
\ ,
\label{g6}
\\
g (y)
& := &
\frac{-\tanh\{(\beta J/2) (1-y^2/2))}{\sqrt{1-y^2/4}} 
\ .
\label{g7}
\end{eqnarray}
Note that $g(y)$
is an analytic function within the entire integration domain
in {Eq.}~(\ref{g6}), i.e. for all $y\in [-\sqrt{2},\sqrt{2}]$.
(The singularities are at $y=\pm 2$.)
Hence, $g(y)$ can be written for all $y\in [-\sqrt{2},\sqrt{2}]$
as a convergent  power series of the form
\begin{eqnarray}
g(y)= \sum_{n=0}^\infty b_n\, y^{2n}
\label{g8}
\end{eqnarray}
with real coefficients 
\begin{eqnarray}
b_0=-\tanh(\beta J/2)\, ,\ 
b_1=\frac{b_0}{8}+\frac{\beta J}{4\cosh(\beta J/2)} \, ,\  ...
\ \ \ \ \ \ \ 
\label{g9}
\end{eqnarray}
Accordingly, {Eq.}~(\ref{g6}) can be rewritten as
\begin{eqnarray}
K (t)
& = &
\sum_{n=0}^\infty b_n \, c_n(t)
\ ,
\label{g10}
\\
c_n(t)
& := &
\int_{-\sqrt{2}}^{\sqrt{2}} dy\,
y^{2n}
\exp\{it(1- y^2/2)\}
\ .
\label{g11}
\end{eqnarray}
Upon partial integration
one thus obtains the recursion relation
\begin{eqnarray}
c_n(t)=\frac{i}{t}\left( 2^{n+1/2} - (2n-1) c_{n-1}(t)\right)
\ .
\label{g12}
\end{eqnarray}
Moreover, one can infer from {Eq.}~(\ref{g11}) that
\begin{eqnarray}
c_0(t) & = & 
\sqrt{\frac{2}{t}}
\,e^{it}\, h(\sqrt{t})
\ ,
\label{g13}
\\
h(t) & := & \int_{-t}^t dx \, e^{-ix^2}
\ .
\label{g14}
\end{eqnarray}

Defining the complex function
\begin{eqnarray}
f(z):=e^{-z^2}
\ ,
\label{g15}
\end{eqnarray}
which is analytical for all $z\in\CC$,
and the path in the complex plane
\begin{eqnarray}
\gamma_1(s):=e^{i\pi/4} s 
\ \  \mbox{with $s\in[-t,t]$,}
\label{g16}
\end{eqnarray}
we can rewrite {Eq.}~(\ref{g14}) as
\begin{eqnarray}
h(t) = e^{-i\pi/4} \int_{\gamma_1} dz \, f(z)
\ .
\label{g17}
\end{eqnarray}
Furthermore, standard arguments of complex analysis imply
\begin{eqnarray} 
\int_{\gamma_1}\! dz  f(z)
= 
\int_{\gamma_2} \! dz  f(z)
+\int_{\gamma_3} \! dz  f(z)
-\int_{\gamma_4} \! dz  f(z)
,
\ \ \ \ \ \ \ \
\label{g18}
\end{eqnarray}
where
\begin{eqnarray}
\gamma_2(s)
& := & s
\ \  \mbox{with $s\in[-t,t]$,}
\label{g19}
\\
\gamma_3(s)
& := & t\, e^{is}
\ \ \mbox{with $s\in[0,\pi/4]$,}
\label{g20}
\\
\gamma_4(s)
& := & t\, e^{is}
\ \ \mbox{with $s\in[\pi,\pi+\pi/4]$.}
\label{g21}
\end{eqnarray}
For large $t$ this implies the approximation
\begin{eqnarray}
\int_{\gamma_2} dz \, f(z) = \sqrt{\pi}
\label{g22}
\end{eqnarray}
up to corrections which are exponentially small in $t$.
Finally, one finds that 
\begin{eqnarray}
& & 
\int_{\gamma_4} \! dz  f(z)
=-\int_{\gamma_3} \! dz  f(z)
\label{g23}
\end{eqnarray}
and that (see Appendix \ref{appH} for the details)
\begin{eqnarray}
\int_{\gamma_3} dz \, f(z) = 
\frac{i}{2t} e^{i(\pi/4 - t^2)} 
\left(1 + i/2t^2 + \ord(t^{-4})\right)
.
\ \ \ \ \ \ 
\label{g24}
\end{eqnarray}
Altogether, we thus can conclude from 
{Eqs.}~(\ref{g17}), (\ref{g18}), (\ref{g22})-(\ref{g24})
 that
\begin{eqnarray}
h(t) = e^{-i\pi/4} \, \sqrt{\pi}
+ \frac{i}{t} e^{-it^2} \left(1 + i/2t^2 + \ord(t^{-4})\right)
\ \ \ \ \ 
\label{g25}
\end{eqnarray}
and with {Eq.}~(\ref{g13}) that
\begin{eqnarray}
c_0 (t) & = & \sqrt{\frac{2\pi}{t}}\,e^{i(t-\pi/4)}
+
i\frac{\sqrt{2}}{t}
-
\frac{1}{\sqrt{2}t^2}
+ 
\ord(t^{-3})
.
\ \ \ \ \ 
\label{g26}
\end{eqnarray}
Furthermore, {Eq.}~(\ref{g12}) yields
\begin{eqnarray}
c_1(t) 
& = &
i\frac{2^{3/2}}{t} 
- 
i \sqrt{\frac{2\pi}{t^3}}\,e^{i(t-\pi/4)}
+
\frac{\sqrt{2}}{t^2}
+
\ord(t^{-3})
\ \ \ \ \ \ \ \ 
\label{g27}
\end{eqnarray}
and (upon iteration of {Eq.}~(\ref{g12}))
\begin{eqnarray}
c_n(t)
& = & 
i\frac{2^{n+1/2}}{t}
- 
\delta_{n,2} \frac{6}{t}\sqrt{\frac{2\pi}{t}}\,e^{i(t-\pi/4)}
\nonumber
\\
& + &
\frac{2n-1}{t^2}2^{n-1/2}
+ \ord(t^{-3})
\ \ \mbox{for $n\geq 2$.}
\ \ \ \ \ \ 
\label{g28}
\end{eqnarray}
Introducing these results into {Eq.}~(\ref{g10}) yields
\begin{eqnarray}
K(t)
& = &
\sqrt{\frac{2\pi}{t}}
\!\left( \!
b_0-i\frac{b_1}{t} 
\! \right)\!
e^{i(t-\pi/4)} 
+ R(t) + 
\ord (t^{-3})
,
\ \ \ \ \ \ \ 
\label{g29} 
\\
R(t)
& := &
i \frac{\sqrt{2}}{t} \sum_{n=0}^\infty b_n 2^n
+
\frac{1}{\sqrt{2}t^2}
\sum_{n=0}^\infty b_n (2n-1) 2^n
\ .
\label{g30}
\end{eqnarray}
Observing {Eq.}~(\ref{g8}), the first sum in {Eq.}~(\ref{g30}) can be identified 
with $g(\sqrt{2})$, and hence with zero according to {Eq.}~(\ref{g7}).
Likewise, in the second sum in {Eq.}~(\ref{g30}) we 
can replace $(2n-1)$ by $2n$, yielding
\begin{eqnarray}
R(t)
=
\frac{2^{3/2}}{t^2}
\sum_{n=0}^\infty b_n n \,2^{n-1}
= 
\frac{2^{3/2}}{t^2} g'(\sqrt{2})
=
\frac{2^{3/2}\beta J}{t^2} 
.
\ \ \ \ \ \  \ \ 
\label{g31}
\end{eqnarray}
By exploiting {Eqs.}~(\ref{g29}) and (\ref{g31}) in (\ref{g5}) we obtain
\begin{eqnarray}
\Im (\tilde B_0(t))
& = &
\sqrt{\frac{2}{\pi t}}  b_0 \sin(t-\pi/4)
\nonumber
\\
& - &
\sqrt{\frac{2}{\pi t^3}}  b_1 \cos (t-\pi/4)
+\ord (t^{-3})
\ .
\ \ 
\label{g32}
\end{eqnarray} 
Introducing this result and
the large-$t$ asymptotics for $J_0(t)$ from 
{Eq.}~(\ref{f8}) into {Eq.}~(\ref{68}) yields
Eq.~(\ref{69}) in the main text.

\section{Derivation of Eq.~(\ref{g24})}
\label{appH}

Exploiting {Eqs.}~(\ref{g15}) and (\ref{g20}) yields
\begin{eqnarray}
\int_{\gamma_3} dz\, f(z)
=
\int_0^{\pi/4} ds\, it \,e^{is}\, e^{-t^2 e^{2is}}
\ .
\label{h1}
\end{eqnarray}
Employing $x:=\pi/2-2s$ as new integration variable one finds that
\begin{eqnarray}
& & 
\int_{\gamma_3} dz\, f(z)
= \frac{i}{2t} e^{i(\pi/4 - t^2)} 
I(t)
\ ,
\label{h2}
\\
& & I(t)
:=
t^2 \int_0^{\pi/2} dx\, e^{-ix/2}\, e^{-t^2(e^{i(\pi/2-x)}-i)}
\nonumber
\\
& & 
=
t^2 \int_0^{\pi/2} dx\, e^{-ix/2}\, e^{-i t^2(\cos(x) -1)}\, e^{-t^2\sin(x)}
\ .
\label{h3}
\end{eqnarray}
For large values of $t$, the last factor implies that
the integral is dominated by small values of $x$.
Rewriting $\cos(x)$ and $\sin(x)$ as power series
and employing $y:=t^2 x$ as new integration variable 
then yields $I (t) = 1 + i/2t^2+\ord(t^{-4})$.
Together with {Eq.}~(\ref{h2}) we thus recover {Eq.}~(\ref{g24}).

\section{Counterpart of Eq.~(\ref{84}) for $V=s_\nu^y$}
\label{appI}

Our objective is to show that the counterpart of the approximation 
(\ref{84}) for perturbations $V=s_\nu^y$ instead of 
$V=s_\nu^x$ is given by
\begin{eqnarray}
\langle s_\alpha^x \rangle_t - \langle s_\alpha^x\rangle _{\rm th}
& = & 
\delta_{\nu \alpha} \frac{g\beta}{4} 
\sin({\hh t}) 
e^{-\omega^2 t^2/4}
\ .
\label{i1}
\end{eqnarray}

According to the assumptions above {Eq.}~(\ref{84}), only the
first summand with $k=0$ has to be kept in {Eq.}~(\ref{13}). 
In view of {Eq.}~(\ref{12}), the approximation (\ref{84})
is therefore equivalent to
\begin{eqnarray}
C_x(t) & = & f(t)  \cos(\hh t)
\label{i2}
\end{eqnarray}
with the abbreviations
\begin{eqnarray}
C_x(t) & := & C_{\! s_\nu^x \!s_\alpha^x}(t)
\ ,
\label{i3}
\\
f(t) & := & \delta_{\nu \alpha} \frac{g\beta}{4} e^{-\omega^2 t^2/4}
\ ,
\label{i4}
\end{eqnarray}
{where} the dependence of $C_x(t)$ and $f(t)$ on $\nu$ and $\alpha$
has been omitted.
Likewise, {Eq.}~(\ref{i1}) is equivalent to
\begin{eqnarray}
C_{\! s_\nu^y \!s_\alpha^x}(t) = f(t) \sin(\hh t)
\ .
\label{i5}
\end{eqnarray}
For symmetry {reasons} (see also above Eq.~(\ref{76})),
this relation is in turn equivalent to
\begin{eqnarray}
C_y(t) & = & - f(t)  \sin(\hh t)
\ ,
\label{i6}
\end{eqnarray}
where
\begin{eqnarray}
C_y(t) & := & C_{\! s_\nu^x \!s_\alpha^y}(t)
\ .
\label{i7}
\end{eqnarray}
In summary, 
{Eq.}~(\ref{i2}) is considered as given and 
{Eq.}~(\ref{i6}) as to be shown.

Similarly as in {Eq.}~(\ref{6}), we define 
\begin{eqnarray}
H' & := & H + G S^z
\ ,
\label{i8}
\end{eqnarray}
where 
\begin{eqnarray}
S^z & := & \sum_{l=1}^L s_l^z
\ .
\label{i9}
\end{eqnarray}
In other words, $H'$ is given by the right-hand side of 
{Eq.}~(\ref{6}) with $G=0$.
In the same way, $\rho_{can}'$ is obtained by replacing
$H$ by $H'$ in {Eq.}~(\ref{4}), and analogously for $A'(t)$
in {Eq.}~(\ref{10}), $A'_{\rm th}$ in {Eq.}~(\ref{11}), and
$C'_{\! V\!\!A}(t)$ in {Eq.}~(\ref{12}).
Finally, {Eq.}~(\ref{b43}) together with {Eqs.}~(\ref{10})-(\ref{12})
implies that {Eqs.}~(\ref{i2}), (\ref{i3}) can be rewritten as
\begin{eqnarray}
C_x(t)= \tr\!\left\{\rho_{can} s_\nu^x e^{iHt} s_\alpha^x e^{-iHt}\right\}
& = & f(t)  \cos(\hh t)
\ . \ \ \ 
\label{i10}
\end{eqnarray}
Similarly, it follows by recalling that $H'$ 
amounts to the case $G=0$ (see below {Eq.}~(\ref{i9})) that
\begin{eqnarray}
\tr\!\left\{\rho'_{can} s_\nu^x e^{iH't} s_\alpha^x e^{-iH't}\right\}
& = & f(t) 
\ ,
\label{i11}
\end{eqnarray}
while {Eq.}~(\ref{i7}) can now be rewritten as
\begin{eqnarray}
C_y(t)= \tr\!\left\{\rho_{can} s_\nu^x e^{iHt} s_\alpha^y e^{-iHt}\right\}
\ .
\label{i12}
\end{eqnarray}

As mentioned above {Eq.}~(\ref{7}) and below {Eq.}~(\ref{a1}), 
$S^z$ commutes with $H$, implying together with {Eq.}~(\ref{i8}) that
$e^{iH't}=e^{iHt}e^{iGS^zt}$. Moreover, all summands
$s_l^z$ in {Eq.}~(\ref{i9}) commute with each other,
and all $s_l^z$ with $l\not=\alpha$ commute
with $s_\alpha^x$, as detailed above {Eq.}~(\ref{a2}).
We thus can conclude that
\begin{eqnarray}
e^{iH't} s_\alpha^x e^{-iH't} 
& = & e^{iHt} B_t e^{-iHt}
\ ,
\label{i13}
\\
B_t & :=& e^{i G s_\alpha^z t} s_\alpha^x e^{-i G s_\alpha^z t}
\ .
\label{i14}
\end{eqnarray}
Working in terms of Pauli matrices (see above Eq.~(\ref{a2})),
a straightforward calculation then yields
\begin{eqnarray}
B_t = s_\alpha^x  \cos(Gt) - s_\alpha^y \sin(Gt) 
\ .
\label{i15}
\end{eqnarray}

Let us now assume that $\rho'_{can}$ in {Eq.}~(\ref{i11}) can be
approximately replaced by $\rho_{can}$.
One thus can conclude from {Eqs.}~(\ref{i10})-(\ref{i15})
that
\begin{eqnarray}
f(t) \cos^2(Gt) - \sin(Gt) \, C_y(t) = f(t)
\label{i16}
\end{eqnarray}
One readily sees that {Eq.}~(\ref{i6}) is the unique solution of this
equation for all $t$ with $\sin(Gt)\not= 0$.
For the remaining $t$'s, the same result (\ref{i6})
follows under the physically obvious extra assumption that 
$C_y(t)$ must be a continuous function of $t$.
In other words, in order to accomplish our goal 
of validating {Eq.}~(\ref{i6}), we are left to justify our 
approximation of $\rho'_{can}$ in {Eq.}~(\ref{i11}) by $\rho_{can}$.
Generally speaking, it seems reasonable to expect
that this may actually be impossible.
On the other hand, for $\beta\to 0$ it is quite obvious
that $\rho'_{can}=\rho_{can}$, hence our approximation
can be justified at least for sufficiently small $\beta$.
Since the same has 
already been taken for granted
above Eq.~(\ref{84}), 
and therefore also in {Eqs.}~(\ref{i1}) and (\ref{i6}),
this argument is sufficient 
for our present purposes.




\begin{thebibliography}{70}
\bibitem{foot2}
The ``$+1$'' on the left-hand side of {Eq.}~(\ref{55}) is needed to guarantee
$L\gg 1$ even if $\nu=\alpha$.
\bibitem{foot4}
Viewing a ``model'' as being defined by a 
Hamiltonian $H$ and an initial state $\rho(0)$, 
a statistical ensemble of models is usually generated by 
randomizing either $H$ or $\rho(0)$ in some 
suitable way.
\end{thebibliography}
\end{document}